\documentclass[twocolumn,twoside,slac_two]{revtex4}
\usepackage{graphicx}
\usepackage{fancyhdr}
\newcommand{\etal}{{\it et al.}}

\begin{document}
%fpcp06_311
%Title of paper
\title{Hadronic Charm Decays and {\boldmath $D$} Mixing}

% Repeat the \author .. \affiliation  etc. as needed
%
% \affiliation command applies to all authors since the last
% \affiliation command. The \affiliation command should follow the
% other information

\author{Sheldon Stone}
\affiliation{Department of Physics, Syracuse University, Syracuse,
New York 13244-1130}

\begin{abstract}
I discuss new results on absolute branching ratios of charm mesons
into specific exclusive final states, Cabibbo suppressed decay
rates, inclusive decays to $s\overline{s}$ mesons, limits on
$D^0-\overline{D}^0$ mixing, CP violation and T violation.
Preliminary results from CLEO-c now dominate the world average
absolute branching fractions. For the most important normalization
modes involving $D^0$ and $D^+$, the averages are ${\cal {B}}(D^0\to
K^-\pi^+)= (3.87\pm 0.06)\%,{\rm~ and~} {\cal {B}}(D^+\to
K^-\pi^+\pi^+)= (9.12\pm 0.19)\%.$ For the $D_S^+$ CLEO-c measures
${\cal {B}}(D_S^+\to K^-K^+\pi^+)=(4.54^{+0.44}_{-0.42}\pm 0.25)\%$.
Using this rate, I derive an effective branching ratio
${\cal{B}}^{\rm eff}(D_S^+\to\phi\pi^+)=(3.49\pm 0.39)\%$, that is
appropriate for use in extracting other branching fractions that
have often been measured relative to this mode. This number is
compared with other determinations.
\end{abstract}

%\maketitle must follow title, authors, abstract
\maketitle

\thispagestyle{fancy}

% body of paper here - Use proper section commands
% References should be done using the \cite, \ref, and \label commands
% Put \label in argument of \section for cross-referencing
%\section{\label{}}

\section{Introduction}

Studies of charm decays are pursued for several different reasons.
First of all, there is the possibility of directly observing new
physics beyond the Standard Model (SM), since the effects of CP
violation due to SM processes is highly suppressed allowing new
physics contributions to be more easily seen than in $b$ decays
where the SM processes typically have large effects \cite{Bigi}.
$D^0-\overline{D}^0$ mixing also is interesting because it could
come from either SM or New Physics (NP) processes, and could teach
us interesting lessons.

Another important reason for detailed charm studies is that most
$b$'s, $\sim$99\%, decay into charm, so knowledge about charm decays
is particularly useful for $b$ decay studies. Especially interesting
are absolute branching ratios, resonant substructures in multi-body
decays, phases on Dalitz plots, etc.. Other heavier objects such as
top quarks decay into $b$ quarks and Higgs particles may decay with
large rates to $b\overline{b}$, again making charm studies
important. Furthermore, charm can teach us a great deal about strong
interactions, especially decay constants and final state
interactions.

\section{Experimental Techniques}
Charm has been studied at $e^+e^-$ colliders at threshold, first by
the Mark III collaboration and more recently by BES and CLEO-c, at
higher $e^+e^-$ energies, and at fixed target and hadron collider
experiments \cite{Artuso}.

The detection techniques are rather different at threshold than in
other experiments. The $\psi(3770)$ resonance decays into
$D\overline{D}$; the world average cross-section is 3.72$\pm$0.09 nb
for $D^0\overline{D}^0$ production and 2.82$\pm$0.09 nb for $D^+D^-$
production \cite{Artuso}. $D_S^+$ production is studied at 4170 MeV,
where the cross-section for $D_S^{*+}D_S^-$+$D_S^{+}D_S^{*-}$ is
$\sim$1 nb \cite{Poling}. The underlying light quark ``continuum"
background is about 14 nb. The relatively large cross-sections,
relatively large branching ratios and sufficient luminosities, allow
experiments to fully reconstruct one $D$ as a ``tag." Since the
charge and flavor of the tag is then uniquely determined, the rest
of the event can be examined for characteristics of the other
``known" particle. To measure absolute branching ratios, for example
at the $\psi(3770)$, the rest of the event is fully reconstructed,
as well as the tag.

At the $\psi(3770)$ $D$ meson final states are reconstructed by
first evaluating the difference in the energy, $\Delta E$, of the
decay products with the beam energy. Candidates with $\Delta E$
consistent with zero are selected and then the $D$
beam-constrained mass is evaluated,
\begin{equation}
m_{\mathrm{BC}}=\sqrt{E_{\mathrm{beam}}^2-(\sum_i\overrightarrow{p}_{\!i})^2},\label{eq:mBC}
\end{equation}
where $i$ runs over all the final state particles.

\begin{figure}[th]
%\centering
\includegraphics[width=80mm]{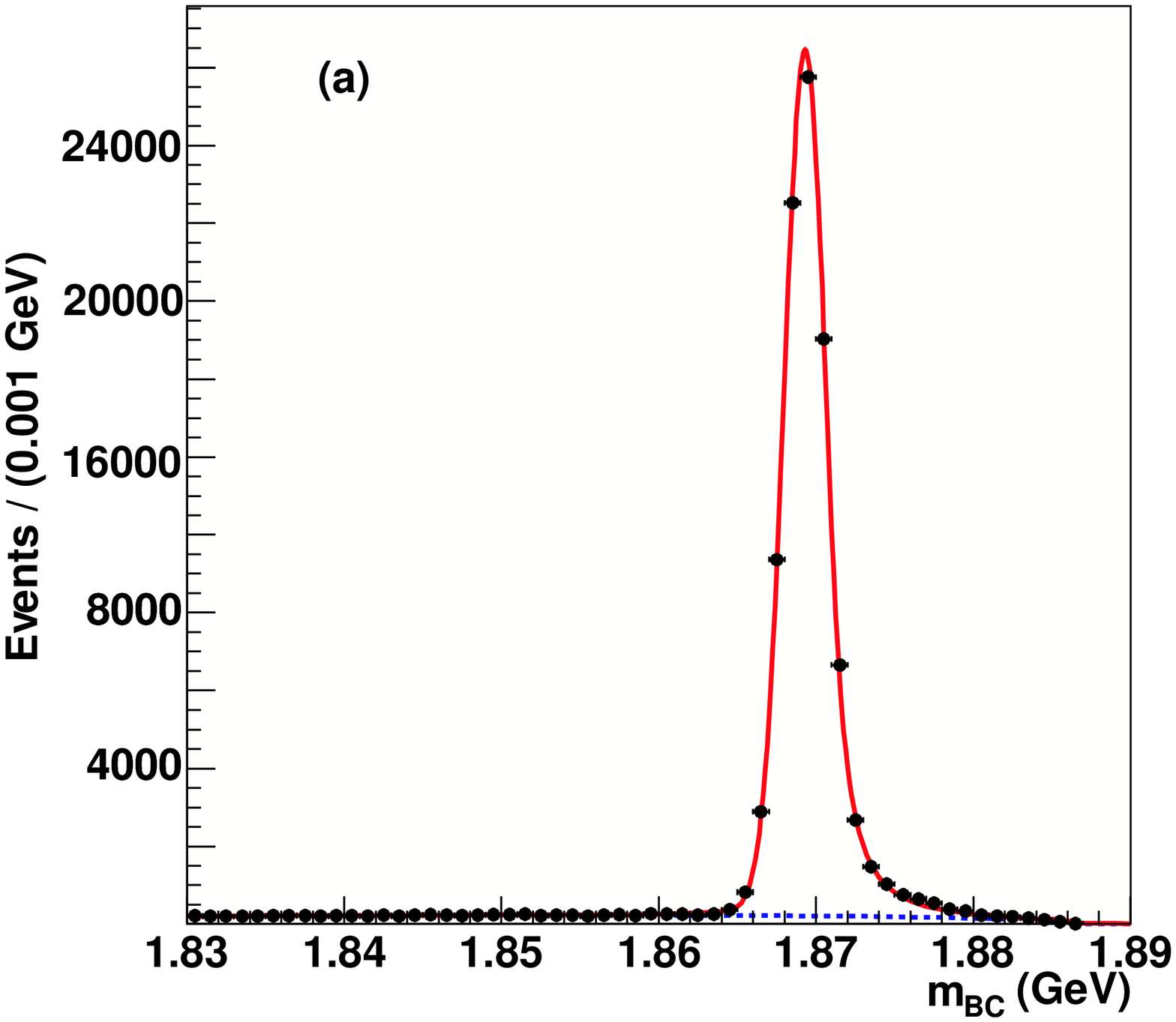}
\includegraphics[width=78mm]{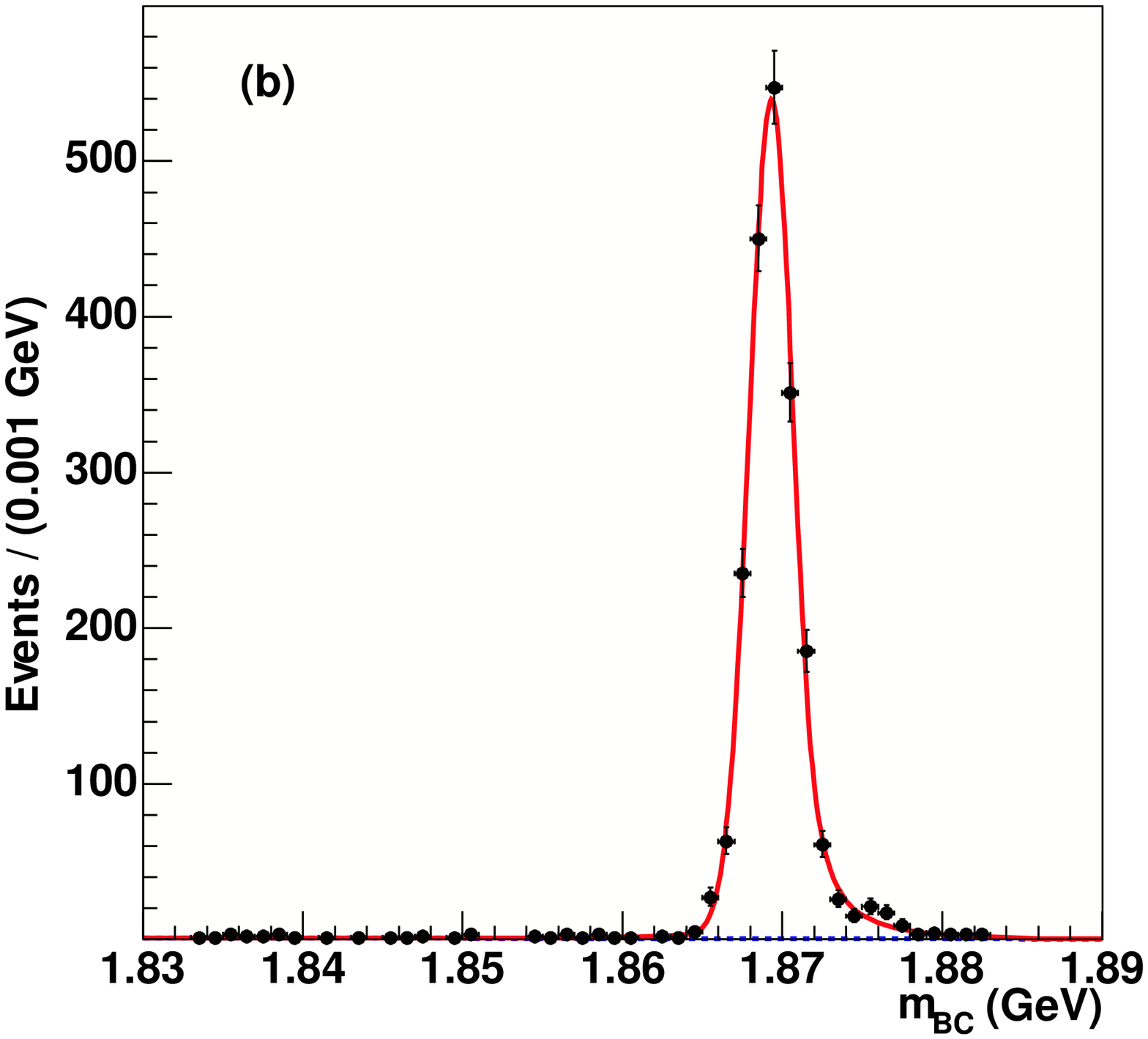}
\caption{(a) The $m_{\mathrm{BC}}$ distributions for candidates from
either $D^+\to K^-\pi^+\pi^+$ or $D^-\to K^+\pi^-\pi^-$ modes. (b)
The $m_{\mathrm{BC}}$ distribution for candidates for candidates
from $D^+\to K^-\pi^+\pi^+$ and $D^-\to K^+\pi^-\pi^-$ modes. The
solid curves are a fits to the signals plus the backgrounds, that
are indicated by the dashed shapes. The signals are asymmetric due
to radiation of the electron beams.} \label{cleo-double}
\end{figure}

Examples of single and double reconstruction are presented in
Fig.~\ref{cleo-double}(a) that shows the $m_{BC}$ distribution for a
$D^+\to K^-\pi^+\pi^+$ or $D^-\to K^+\pi^-\pi^-$ final states. These
``single tags" show a large signal and a very small background.
Fig.~\ref{cleo-double}(b) shows a ``double" tag sample where both
$D^+$ and $D^-$ candidates in the same event are reconstructed.

Other experiments make use of the both the approximately picosecond
lifetimes of charm to identify detached vertices, and the decay
$D^{*}\to \pi D$, which also serves as a flavor tag in the case of
$D^{*\pm}\to \pi^{\pm} D^0$ transitions.

\section{Absolute Charm Meson Branching Ratios and Other Hadronic Decays}

\subsection{Absolute {\boldmath $D^0$ and $D^+$} Branching Ratios}
In charm meson decays, usually a single branching ratio sets the
scale for determinations of most other rates, that are measured
relative to it. For $D^0$ and $D^+$ these modes are $K^-\pi^+$ and
$K^-\pi^+\pi^+$, respectively.\footnote{In general, mention of a
mode will also imply consideration of the charge-conjugate state.}
CLEO-c, on the other hand uses a different technique where the
branching ratios of several modes are determined simultaneously and
all absolutely. Consider an ensemble of modes $i$, that are both
singly reconstructed and also doubly reconstructed, where all
combinations of modes may be used. I denote the number of observed
single tag charmed particles as $N_i$, anti-charmed particles as
$N_j$, and double tags as $N_{ij}$. They are related to the number
of $D\overline{D}$ events (either charged or neutral) through their
branching ratios ${\cal{B}}_i$ as
\begin{eqnarray}
N_i&=&\epsilon_i {\cal{B}}_i N_{D\overline{D}}\\
N_j&=&\epsilon_j {\cal{B}}_j N_{D\overline{D}}\\
N_{ij}&=&\epsilon_{ij} {\cal{B}}_i{\cal{B}}_j N_{D\overline{D}}~,
\end{eqnarray}
where $\epsilon_i$ and $\epsilon_{ij}$ are the reconstruction
efficiencies in single and double tag events for each mode. (In
practice the differences in each mode between single and double tag
events are small, and $\epsilon_{ij}\approx\epsilon_i\epsilon_j$.)
Solving these equations we find
\begin{eqnarray}
{\cal{B}}_j&=&\frac{N_{ij}}{N_i}\frac{\epsilon_i}{\epsilon_{ij}} \\
N_{D\overline{D}}&=&\frac{N_iN_j}{N_{ij}}\frac{\epsilon_{ij}}{\epsilon_i\epsilon_j}
\end{eqnarray}

CLEO-c has recently updated their absolute branching ratio
measurements using a 281 pb$^{-1}$ data sample, an approximately 5
times larger data sample than used by them for their previous
publication \cite{cleo-dbr}. The new preliminary results are shown
in Table~\ref{tab:Dbr} \cite{Dbr}. (In this table when two errors
follow a number, the first error is statistical and the second
systematic; this will be true for all results quoted in this paper
unless specifically indicated.)

\begin{table}[ht]
\begin{center}
\caption{$D^0$ and $D^+$ Branching Ratios comparing preliminary
CLEO-c results with the PDG \cite{PDG}.}
\begin{tabular}{|l|c|c|}
\hline \boldmath{$D^0$} \textbf{Decays} & \boldmath{${\cal
B}$}\textbf{\%} \textbf{CLEO-c} & \textbf{PDG}
\\\hline
$K^-\pi^+$ &3.839$\pm$0.035$\pm$0.060&3.91$\pm$0.09\\\hline
$K^-\pi^+\pi^0$ &14.46$\pm$0.12$\pm$0.38&13.2$\pm$1.0\\\hline
$K^-\pi^+\pi^+\pi^-$ &8.29$\pm$0.07$\pm$0.21&7.48$\pm$0.30\\\hline
\hline \boldmath{$D^+$} \textbf{Decays} &&
\\\hline
$K^-\pi^+\pi^+$ &9.11$\pm$0.10$\pm$0.17& 9.2$\pm$0.6\\\hline
$K^-\pi^+\pi^+\pi^0$ &5.95$\pm$0.07$\pm$0.17& 6.5$\pm$1.1\\\hline
$K^0\pi^+$ &3.092$\pm$0.044$\pm$0.074& 2.83$\pm$0.18\\\hline
$K^0\pi^+\pi^0$ &14.40$\pm$0.18$\pm$0.58& 10.7$\pm$2.9\\\hline
$K^0\pi^+\pi^+\pi^-$ &6.366$\pm$0.052$\pm$0.184& 7.1$\pm$1.0\\\hline
$K^-K^+\pi^+$ &0.930$\pm$0.016$\pm$0.029& 0.89$\pm$0.08\\\hline
\end{tabular}
\label{tab:Dbr}
\end{center}
\end{table}

\begin{figure}[htb]
%\centering
\includegraphics[width=80mm]{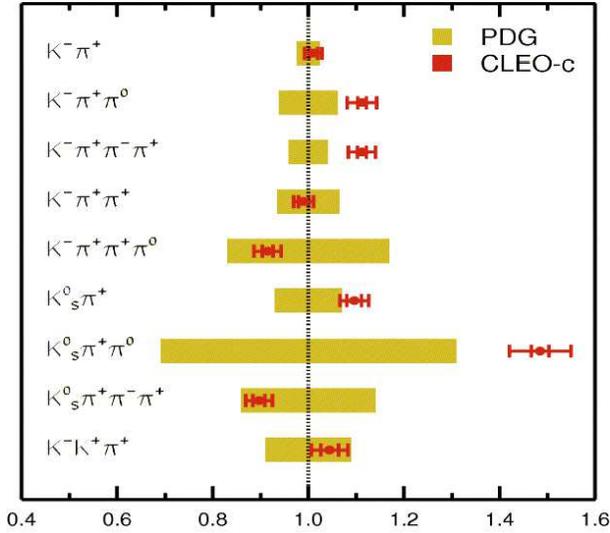}
\caption{Ratio of preliminary CLEO-c results (281 pb$^{-1}$) to
those of the PDG. The smaller error bars on the CLEO-c results
indicate statistical errors and the larger ones systematic errors,
shown linearly.} \label{Dbr2}
\end{figure}

Systematic uncertainties now are the dominant error source in the
CLEO-c results. The largest error, common to all decay modes, is
that on the $\Delta E$ cut that ranges between $\pm$1.0\% and
$\pm$2.5\% depending on the final state. Modes with $K_S$ or $\pi^0$
have additional $\pm$1.1\% and $\pm$2\% errors, respectively. A
comparison with the PDG values excluding all CLEO-c results is shown
pictorially in Fig.~\ref{Dbr2}.

The world average values for the normalizing modes are now dominated
by the CLEO-c results. They are
\begin{eqnarray}
{\cal {B}}(D^0\to K^-\pi^+)&=& (3.87\pm 0.06)\%\\\nonumber {\cal
{B}}(D^+\to K^-\pi^+\pi^+)&=& (9.12\pm 0.19)\%,
\end{eqnarray}
and are now determined with 1.5\% and 2.1\% relative accuracy,
respectively.

\subsection{Analysis Procedures for \boldmath{$D_S^+$} Studies at 4170 MeV}
\begin{figure}[htb]
%\centering
\includegraphics[width=74mm]{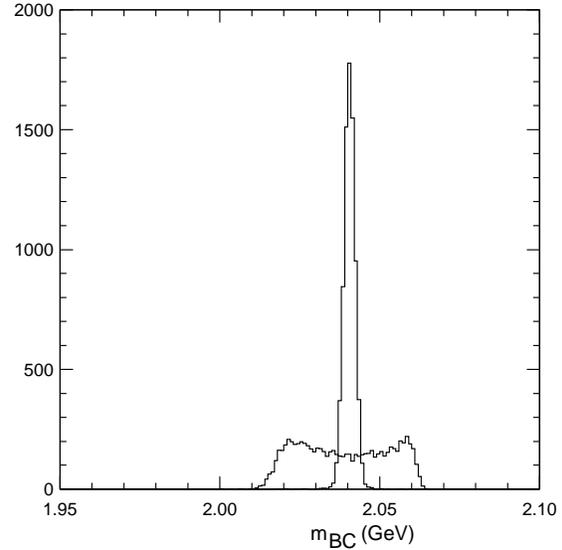}
\vspace{0.44mm}\caption{The beam constrained mass $m_{BC}$ from
Monte Carlo simulation of $e^+e^-\to D_S^+D_S^{*-}$,
$D_S^{\pm}\to\phi\pi^{\pm}$ at 4170 MeV. The narrow peak is from
the $D_S^+$ and the wider one from the $D_S^-$. (The distributions
are not centered at the $D_S^+$ or $D_S^{*+}$ masses, because the
reconstructed particles are assumed to have the energy of the
beam.)} \label{mbc}
\end{figure}

At 4170 MeV $D_S$ mesons are produced mainly in the two processes
$e^+e^-\to D_S^{*+}D_S^-$ or $D_S^{+}D_S^{*-}$. If we do not detect
the photon and reconstruct the $m_{BC}$ distribution making a loose
$\Delta E$ cut using Eq.~\ref{eq:mBC}, we obtain the distribution
from Monte Carlo shown in Fig.~\ref{mbc}. The narrow peak occurs
when the reconstructed $D_S$ does not come from the $D_S^*$ decay.
Thus, the method used so successfully on the $\psi(3770)$ no longer
works as well.

\begin{figure}[ht]
\centering
\includegraphics[width=74mm]{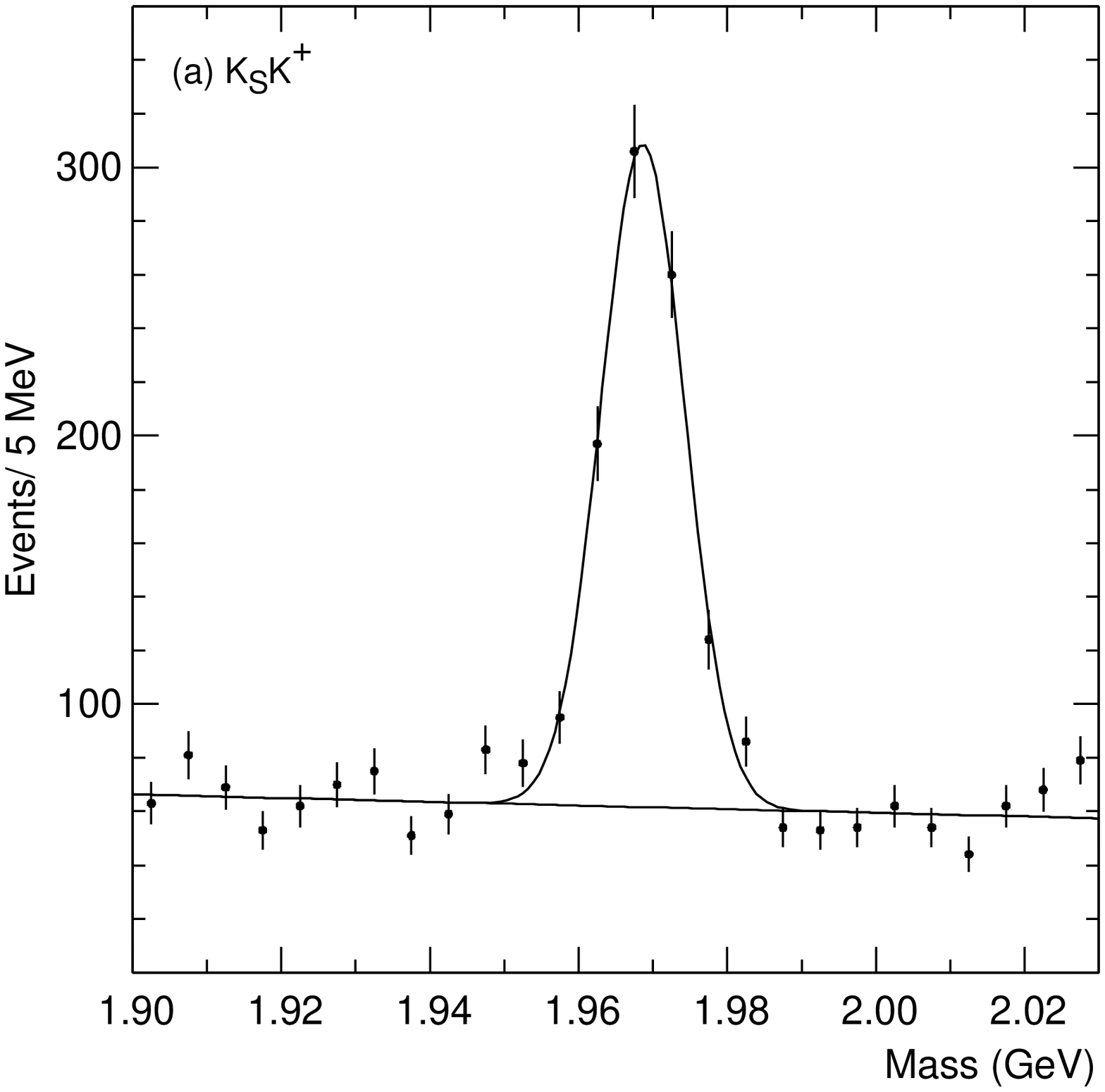}
\includegraphics[width=80mm]{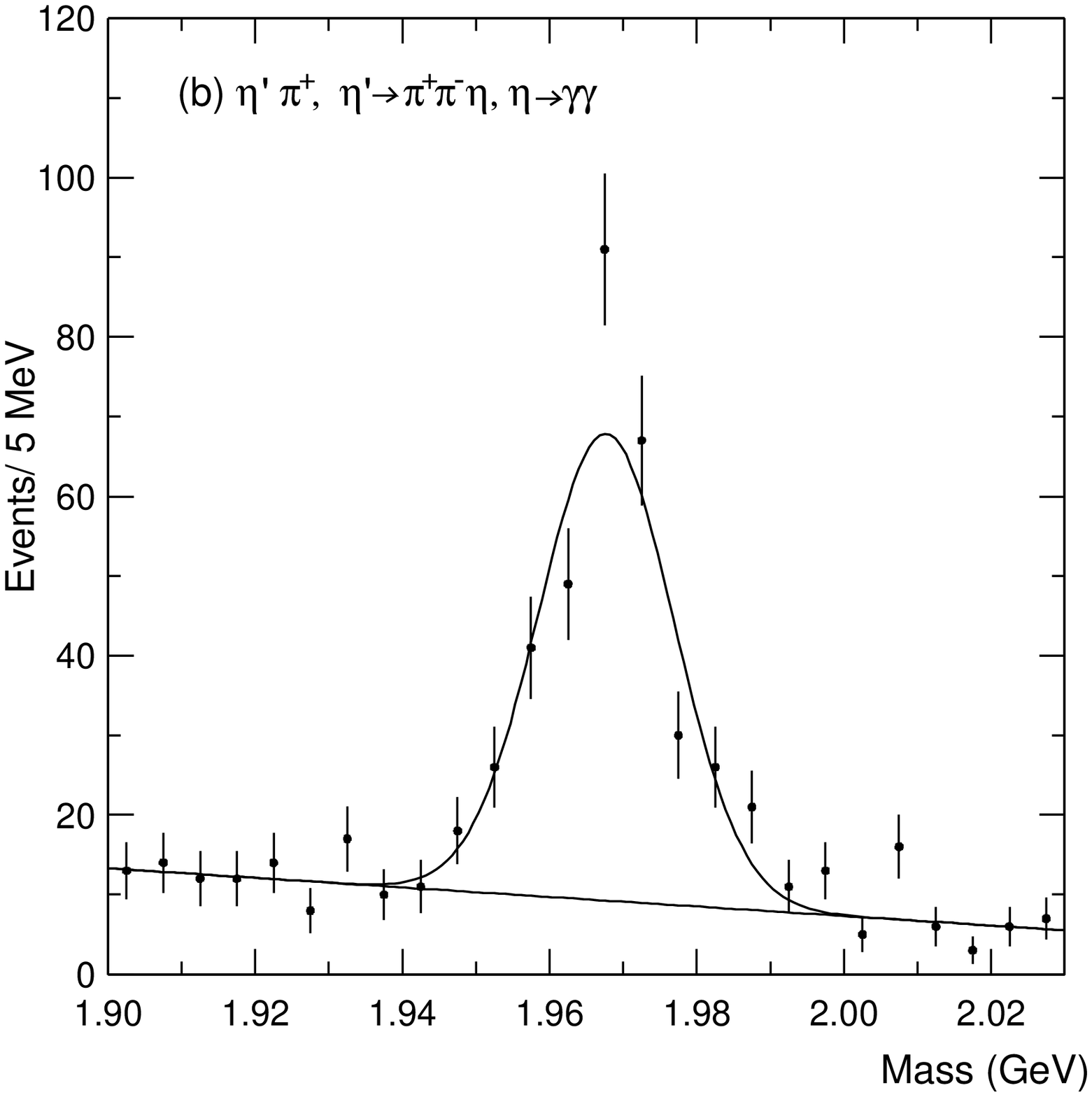}
\caption{Invariant mass of (a) $K_SK^+$ combinations from CLEO-c,
requiring the total energy to be consistent with the beam energy,
and (b) $\eta'\pi^+$ mass combinations, where
$\eta'\to\pi^+\pi^-\eta$, $\eta\to\gamma\gamma$. The curve is a
fit to a Gaussian signal function plus a linear background.
 } \label{Inv-mass}
\end{figure}

The alternative method used by CLEO-c is to require that the energy
of the $D_S$ is reasonably close to the energy expected in
$D_S^*D_S$ events, and then examine the invariant mass. Some such
distributions from data are shown in Fig.~\ref{Inv-mass}. Note that
the resolution in invariant mass is excellent, and the backgrounds
not very large, at least in these modes. These mass distributions
along with others are used for the CLEO-c inclusive $D_S$ analysis
described later.

\begin{figure*}[th]
\centering
\includegraphics[width=135mm]{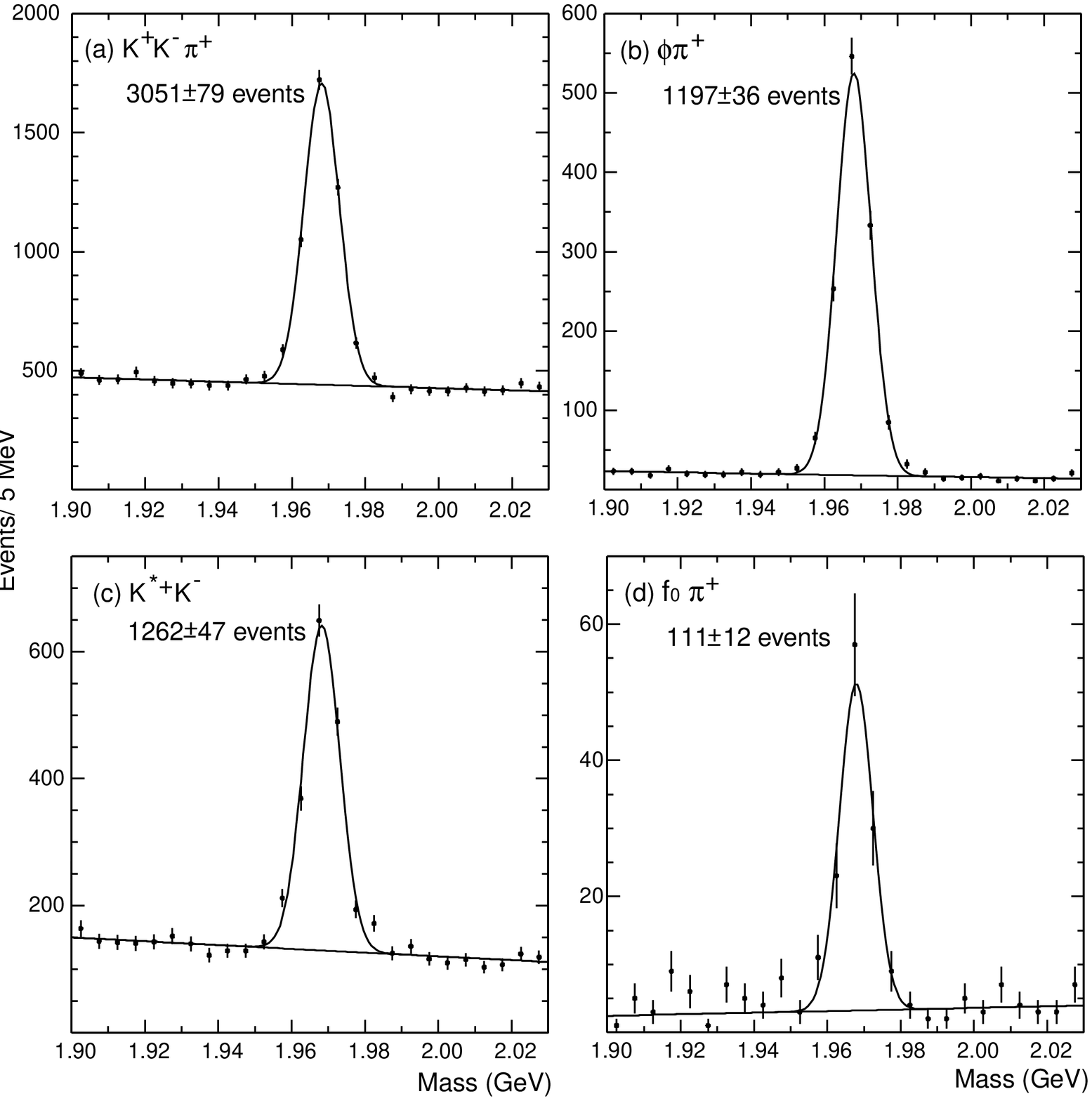}
\caption{Invariant mass of $K^+K^-\pi^+$  combinations from
CLEO-c, requiring the total energy to be consistent with the beam
energy. (a) No requirements on two-body mass. (b) $K^+K^-$ mass in
the $\phi$ region. (c) $K^-\pi^+$ mass in the $K^*(890)$ region.
(d) $K^+K^-$ mass in the $f_0(890)$ region. } \label{KKpi}
\end{figure*}

The $K^+K^-\pi^+$ mode is particularly interesting. I show in
Fig.~\ref{KKpi} the invariant mass distributions (a) independent of
any selections on two-body mass, and for requirements on either
$K^+K^-$ mass (b) within $\pm$20 MeV of the $\phi$ mass, or (c) in
the $f_0(980)$ region from 925 to 1010 MeV in $K^+K^-$ mass, or (d)
in the $K^-\pi^+$ mass region within $\pm$100 MeV of the $K^*(890)$
mass.

\begin{figure}[bht]
%\centering
\includegraphics[width=80mm]{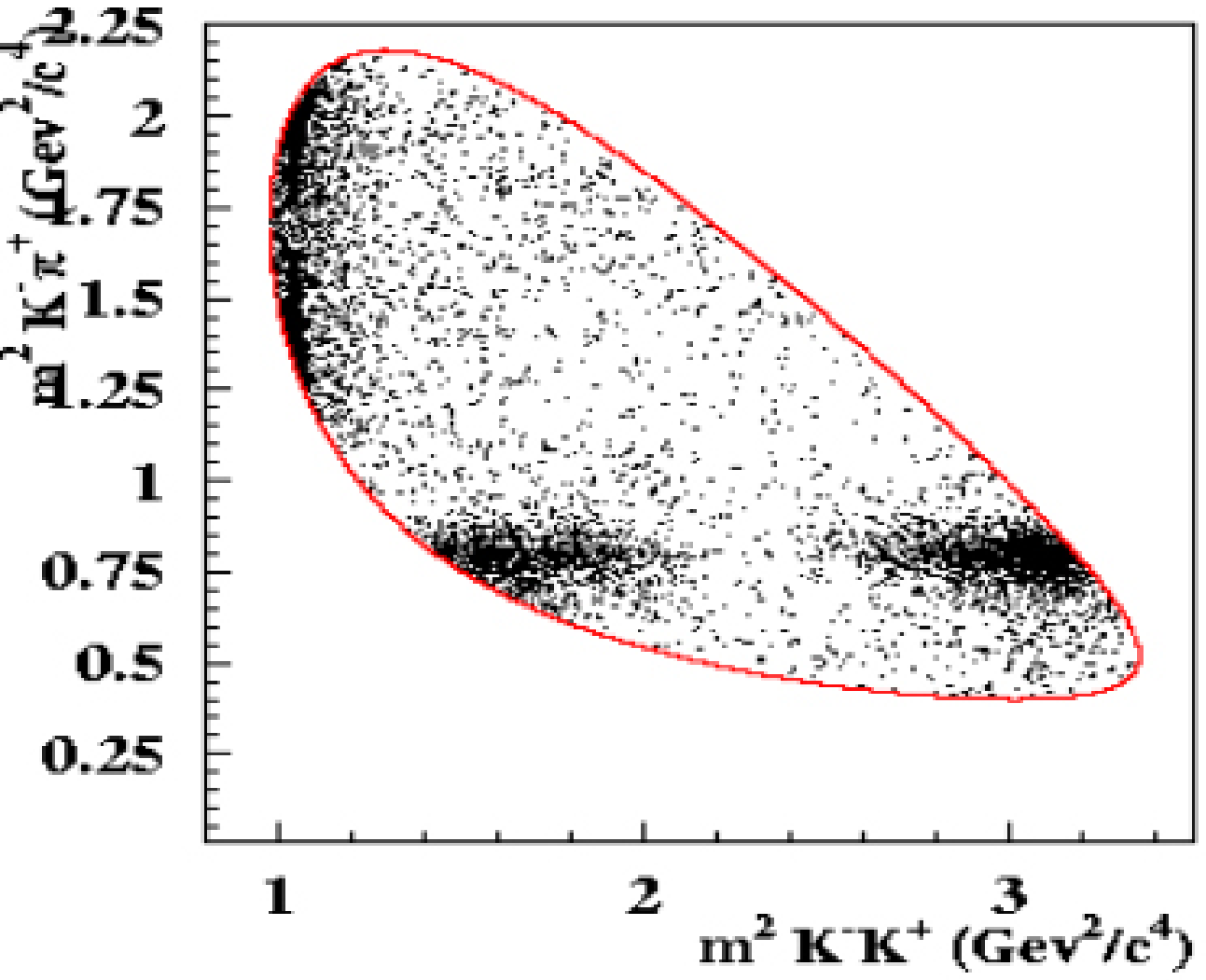}
\caption{Mass-squared of $K^+K^-$ versus mass-squared of
$K^-\pi^+$ in $D_S^+\to K^+K^-\pi^+$ events from Focus.}
\label{focus}
\end{figure}

Both E687 and FOCUS have done Dalitz plot analyses and extracted the
fit fractions and phases for this decay mode. The Dalitz plot from
FOCUS is shown in Fig~\ref{focus}, and the results listed in
Table~\ref{tab:focus} along with the E687 results \cite{Focus,
E687}. The final state is quite complicated. Besides the large
$\phi\pi^+$ and $K^{*0}\pi^+$ components there is a significant
amount of $f_0(980)$ which interferes with the other components as
well as smaller amounts of two other resonances. In particular, the
$f_0$ has an overlap with the $\phi$.

\begin{table}[ht]
\begin{center}
\caption{Final state composition of $D_S^+\to K^+ K^- \pi^+$.}
\begin{tabular}{|l|c|c|c|c|}
\hline & \multicolumn{2}{|c|}{\textbf{Focus}}&\multicolumn{2}{|c|}{\textbf{E687}}\\
& Fraction(\%) & Phase($^\circ$) & Fraction(\%) &
Phase($^\circ$)\\\hline {$K^{*0}(892)$}& 44$\pm$1& 0(fixed)&
48$\pm$5& 0(fixed)\\\hline {$K^*_0(1430)$}& 6$\pm$1&
114$\pm$5&9$\pm$3&152$\pm$40\\\hline {$\phi(1020)$}& 45$\pm$1&
148$\pm$4&40$\pm$3&178$\pm$20\\\hline {$f_0(980)$}& 16$\pm$1&
135$\pm$4&11$\pm$4&159$\pm$22\\\hline {$f_j(1710)$}& 4$\pm$1&
106$\pm$6&3$\pm$2&110$\pm$20\\\hline
\end{tabular}
\label{tab:focus}
\end{center}
\end{table}

\subsection{Absolute {\boldmath $D_S^+$} Branching Ratios}
\begin{figure}[h]
\centering
\includegraphics[width=80mm]{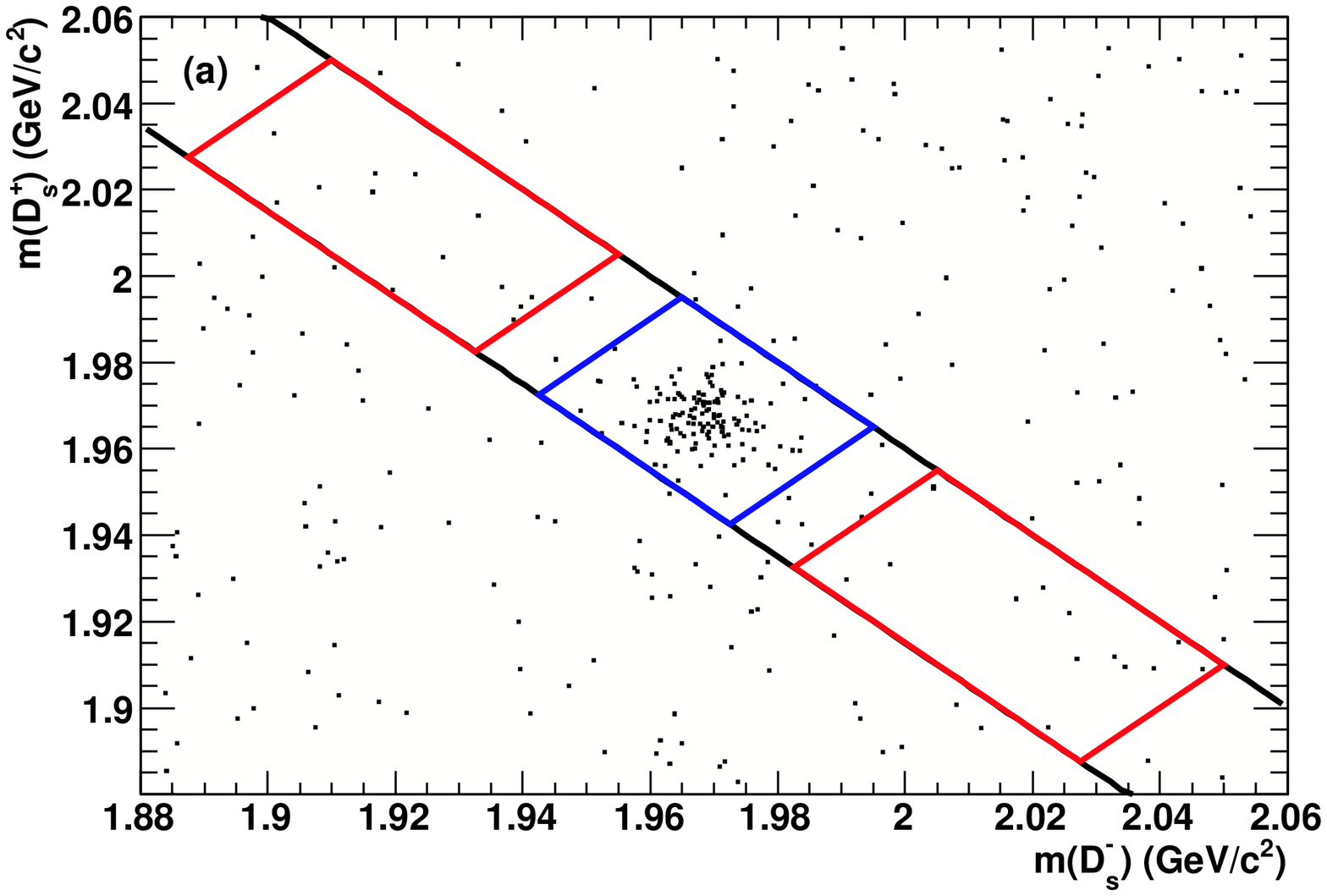}
\includegraphics[width=80mm]{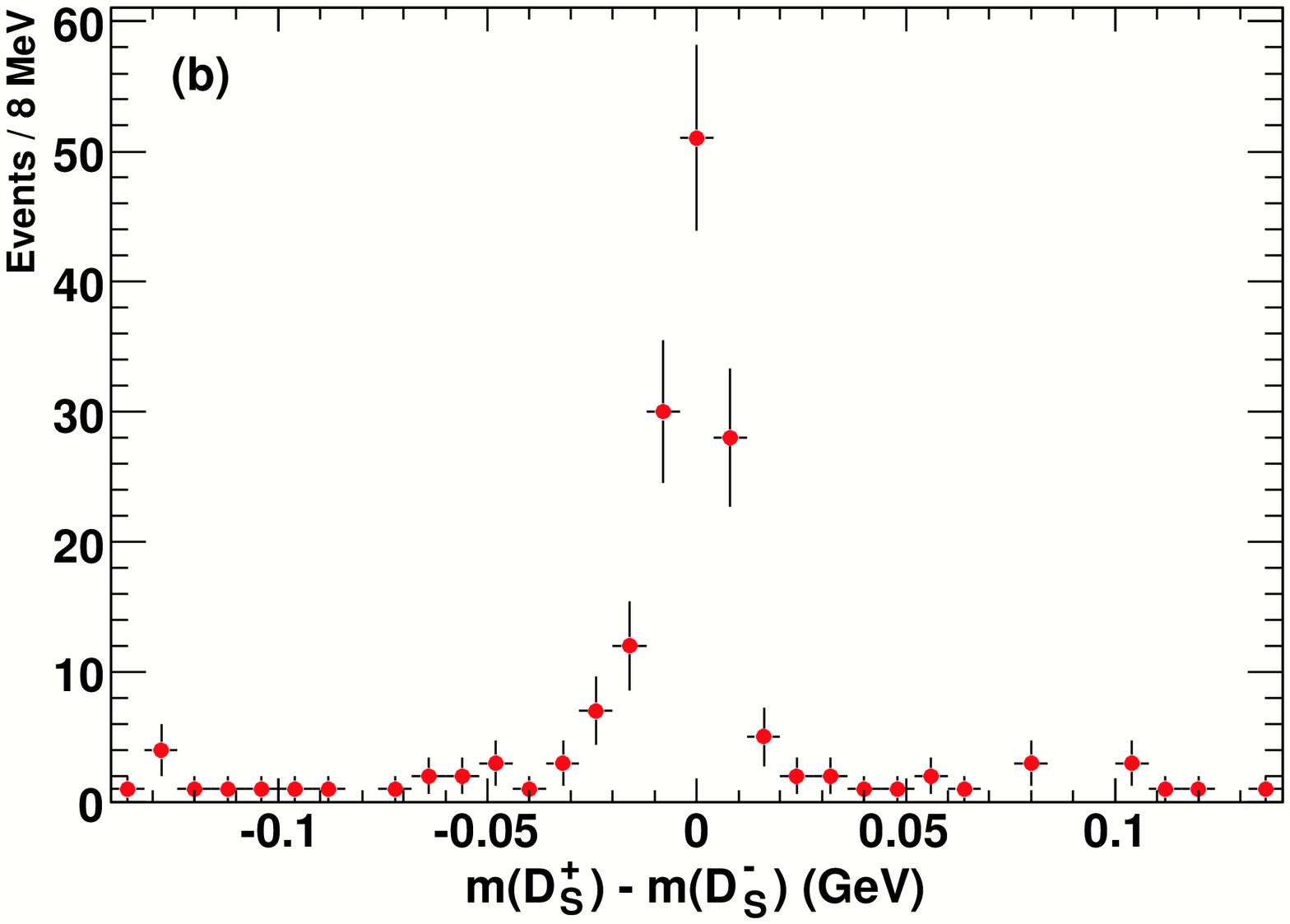}
\caption{Invariant mass distributions in double tag $D_S$ events.
(a) $D_S^+$ versus $D_S^-$; the center rectangle indicates the
signal region and the two others indicate the background samples.
(b) The difference in $D_S^+$ minus $D_S^-$ invariant mass.}
\label{Dm-1d2}
\end{figure}

Measurement of the $D_S$ absolute branching ratios proceeds in a
similar manner as used for the $D^0$ and $D^+$ rates using a 56
pb$^{-1}$ data sample at 4170 MeV and 20 pb$^{-1}$ at nearby
energies. The single tags are reconstructed using invariant mass
with a loose cut on the $D_S$ energy. The photon or $\pi^0$ from the
$D_S^*$ decay is ignored. The modes used and the number of single
and double tags are listed in Table~\ref{tab:Dsevents}. The
invariant mass of the $D_S^-$ candidates versus the $D_S^+$
candidates in a single event are plotted in Fig.~\ref{Dm-1d2}(a),
while the differences in invariant mass are shown in
Fig.~\ref{Dm-1d2}(b). These plots demonstrate very good signal to
background in these modes in double tags.

\begin{table}[ht]
\begin{center}
\caption{The $D_S$ decay modes used in this analysis and the
reconstructed number of single tag and double tag events.}
\begin{tabular}{|l|c|c|c|c|}
\hline \multicolumn{5}{|l|}{\textbf{Single Tags}}\\\hline  &
$K_SK^{\pm}$ & $K^-K^+\pi^{\pm}$ & $K^-K^+\pi^{\pm}\pi^0$ &
$\pi^+\pi^-\pi^{\pm}$\\\hline $D_S^+$& 442$\pm$25 & 1607$\pm$54
&333$\pm$38& 265$\pm$29
\\\hline
$D_S^-$& 346$\pm$23 & 1736$\pm$55 & 376$\pm$38& 274$\pm$28
\\\hline\hline
\multicolumn{5}{|l|}{\textbf{Double Tags}}\\\hline &$K_SK^-$ &
$K^-K^+\pi^-$ & $K^-K^+\pi^-\pi^0$ & $\pi^+\pi^-\pi^-$\\\hline
$K_SK^+$& 4 & 5 &7&3
\\\hline
$K^-K^+\pi^+$ & 2 & 36 & 14 & 13\\\hline
$K^-K^+\pi^+\pi^0$&3&12&5&4\\\hline $\pi^+\pi^-\pi^+$&2 & 8 & 0 &
not used\\\hline
\end{tabular}
\label{tab:Dsevents}
\end{center}
\end{table}

The preliminary CLEO-c absolute branching ratios results are shown
in Table~\ref{tab:Dsbr}. The measurements in the all charged modes
have about an $\pm$11\% error and are already the best in the world.
The accuracy will improve markedly by the end of this summer, as the
data sample will increase by a factor of 2.6 and more modes will be
added.
\begin{table}[ht]
\begin{center}
\caption{The $D_S$ absolute branching fraction results (preliminary)
from CLEO-c using 76 pb$^{-1}$ of data near 4170 MeV.}
\begin{tabular}{|l|c|c|}
\hline \textbf{Mode} & \boldmath{${\cal{B}}$}{\textbf(\%)}
\textbf{CLEO-c} & \boldmath{${\cal{B}}$}\textbf{(\%) PDG}\\\hline
$K^0K^+$& 2.56$^{+0.26}_{-0.24}\pm 0.14$ & 3.6$\pm$1.1\\\hline
$K^-K^+\pi^+$ &4.54$^{+0.44}_{-0.42}\pm 0.25$ &
4.3$\pm$1.2\\\hline $K^-K^+\pi^+\pi^0$& 4.83$^{+0.49}_{-0.46}\pm
0.46$ & -\\\hline $\pi^+\pi^-\pi^+$& 1.02$^{+0.11}_{-0.10}\pm
0.05$ & 1.00$\pm$0.28\\\hline
\end{tabular}
\label{tab:Dsbr}
\end{center}
\end{table}

\subsection{The {\boldmath$D_S^+\to \phi\pi^+$} Absolute Branching Ratio}

Table~\ref{tab:Dsbr} does not include a result for $D_S^+\to
\phi\pi^+$. The reason CLEO-c gives for its absence is that the
definition of what constitutes a $\phi$ is somewhat ambiguous, due
to the interferences in the $K^+K^-\pi^+$ Dalitz plot discussed
above. Most measurements reported for $D_S$ decays in the PDG,
however, are normalized to this rate, making it important to have
an estimate of the effective branching ratio.
 The
value can vary depending on the experimental resolution and what
selection criteria on the $\phi$ mass are applied by different
analyses. The observed shape of the $K^+K^-$ mass distribution in
the $\phi$ region is convolution of the Breit-Wigner natural width
of the $\phi$ ($\Gamma=4.3$ MeV) with the detector resolution,
which often is simply described by a single Gaussian. Since, mass
resolutions of most experiments are somewhat similar, the number
that I derive here may be of some use. The $\phi\pi$ candidate
mass distribution shown in Fig.~\ref{KKpi}(b) uses a $\pm$20 MeV
cut around the $\phi$ mass and is 97\% efficient. Taking the
measured CLEO-c branching fraction for $D_S^+\to K^+K^-\pi^+$
given in Table~\ref{tab:Dsbr}, and the ratio of the number of
events found by fitting Figs.~\ref{KKpi}(b) and \ref{KKpi}(a) for
$D_S$ events, I find ${\cal{B}}^{\rm
eff}(D_S^+\to\phi\pi^+)$=(3.73$\pm$0.42)\%. If the mass cut is
lowered to $\pm$10 MeV with a 91\% efficiency,
\begin{equation}
{\cal{B}}^{\rm eff}(D_S^+\to\phi\pi^+)=(3.49\pm 0.39)\%~,
\end{equation}
where the 7\% reduction is presumably due to elimination of a
large part of the background $f_0(980)$. I choose to quote this
number as the effective branching ratio.

There are several previous measurements of this rate and one
theoretical prediction. CLEO and BaBar use partial and full
reconstruction of the decay $\overline{B}^0\to D^{*+}D_S^{*-}$. To
reduce systematic errors CLEO does the analysis two ways by
partially reconstructing either the $D_S^{*-}$ decay or the $D^{*+}$
decay. They determine
$\frac{{\cal{B}}(D_S^+\to\phi\pi^+)}{{\cal{B}}(D^0\to
K^-\pi^+)}=0.92\pm0.20\pm0.11$ \cite{CLEO-DS1}. Using the $D^0$
branching ratio from Table~\ref{tab:Dbr} results in ${\cal{B}}^{\rm
eff}(D_S^+\to\phi\pi^+)$=(3.5$\pm$0.8$\pm$0.4)\%.

BaBar, on the other hand, compares the partial reconstruction of
the $D_S^{*-}\to \gamma X$ decays with the fully reconstructed
decay $D_S^{*-}\to \gamma D_S^-$, $D_S^-\to\phi\pi^-$
\cite{BaBar-DS}. They find
${\cal{B}}(D_S^+\to\phi\pi^+)$=(4.8$\pm$0.5$\pm$0.4)\%.

At this conference Marsiske presented another result from BaBar.
Here they use full and partial reconstruction of $\overline{B}^0\to
D^{*+}D_{SJ}^{-}(2460)$ where the $D_{SJ}^{-}(2460)\to D_S^+\gamma$
or $D_S^{*+}\pi^0$. They find ${\cal{B}}^{\rm
eff}(D_S^+\to\phi\pi^+)$=(4.8$\pm$0.4$\pm$0.5)\% \cite{Marsiske},
almost identical to their previous result.

The BaBar values of 4.8\%, however, are quite large and almost
incompatible with a previous limit. Muheim and Stone published a
theoretical prediction and an experimental upper limit of 4.8\% at
90\% C. L., based on summing all the measured modes, most of which
were measured with respect to $\phi\pi^+$ \cite{Muheim}. Updating
the limit using current ratios of branching fractions it becomes a
slightly less restrictive 5.2\% at 90\% C. L. (Their prediction for
${\cal{B}}^{\rm eff}(D_S^+\to\phi\pi^+)$ is (3.6$\pm$0.6)\%.)
Finally there is a result from BES based on 2 events of
($3.9^{+5.1+1.8}_{-1.9-1.1})$\% \cite{BES}.

I choose not to make an average of these values as there is more
evidence that shows the BaBar results are too large. This will be
discussed in the next section.

\subsection{Inclusive Charm Meson Decays to \boldmath{$s\overline{s}$} Mesons}
CLEO-c has investigated the decays of charm mesons into lighter
particles that have large $s\overline{s}$ quark content. These
include the $\eta$, $\eta'$ and $\phi$. These rates are determined
using events where one $D$ is fully reconstructed and the other
decays into the meson of choice. Their preliminary results are
listed in Table~\ref{tab:Dinc}.

\begin{table}[ht]
\begin{center}
\caption{$D^0$, $D^+$ and $D_S^+$ inclusive branching ratios into
$\eta$, $\eta'$ and $\phi$ mesons. The $D$ meson decays used 281
pb$^{-1}$ of data at the $\psi(3770)$, while the $D_S$ decays were
measured using 71 pb$^{-1}$ at or near 4170 MeV.}
\begin{tabular}{|l|c|c|c|}
\hline &\boldmath{$\eta$}\textbf{(\%)} &
\boldmath{$\eta'$}\textbf{(\%)}
&\boldmath{$\phi$}\textbf{(\%)}\\\hline
 $D^0$& $9.4\pm 0.4\pm 0.6$ & $2.6\pm 0.2\pm 0.2$ & $1.0\pm 0.1\pm 0.1$\\\hline
 $D^+$& $5.7\pm 0.5\pm 0.5$ & $1.0\pm 0.2\pm 0.1$ & $1.0\pm 0.1\pm 0.2$\\\hline
 $D_S^+$& $32.0\pm 5.6\pm 4.7$ & $11.9\pm 3.3\pm 1.2$ & $15.1\pm 2.1\pm 1.5$
\\\hline
\end{tabular}
\label{tab:Dinc}
\end{center}
\end{table}

We see that $\phi$ and $\eta'$ mesons are relatively rare in $D^0$
and $D^+$ decays, while they are relative prolific in $D_S$
decays. About 60\% of $D_S$ decays have one of these mesons. The
$\eta$ is produced in significant amounts in $D^0$ and $D^+$
decays and has a large 32\% rate in $D_S$ decays. These results
should be useful for hadron collider experiments that use $B_S\to
D_S X$ decay modes.

\begin{figure*}[ht]
\centering
\includegraphics[width=145mm]{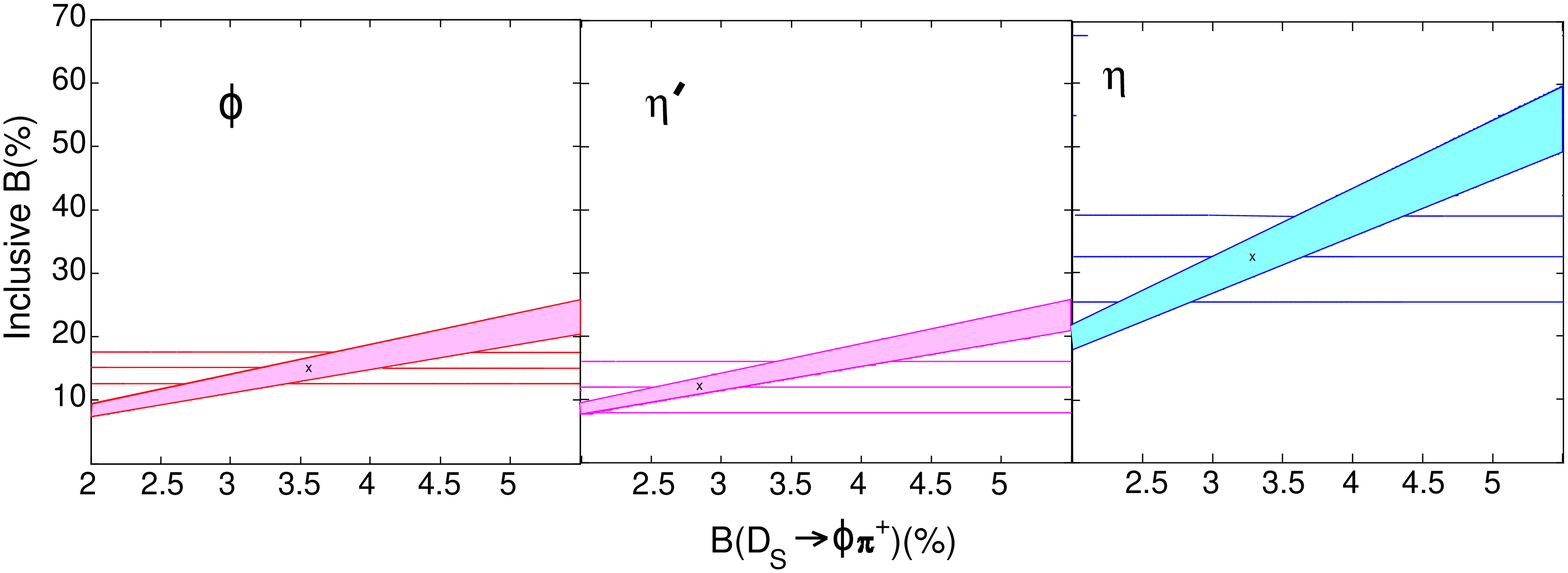}
\caption{The CLEO-c measurements of the inclusive yields of
$\phi$, $\eta'$ and $\eta$ mesons (horizontal lines) with their
$\pm1 \sigma$ errors. Also shown are bands that represent
${\cal{B}}_{\rm SUM}(D_S^+\to \phi X$, $\eta' X$, or $\eta$ X).
These  edges of these bands represent the central value $\pm
1\sigma$. The x indicates the most likely exclusive branching
ratio assuming all modes have been found.}\label{inclu}
\end{figure*}

We can also use these results to check ${\cal{B}}^{\rm
eff}(D_S^+\to\phi\pi^+)$. Actually three independent checks are
possible using each one of these particles. Let us start with the
$\phi$. I simply count the branching fraction for the decays that
include a $\phi$ meson. These include $\phi\pi^+\pi^0$,
$\phi\pi^+\pi^+\pi^-$, $\phi\ell^+\nu$ and $\phi\pi^+$. All of these
modes have been measured with respect to $\phi\pi^+$. Summing them
relates the inclusive $\phi$ yield from the already measured modes
to ${\cal{B}}^{\rm eff}(D_S^+\to\phi\pi^+)$ as
\begin{equation}
{\cal{B}}_{\rm SUM}(D_S^+\to \phi X)=(4.2\pm 0.5){\cal{B}}^{\rm
eff}(D_S^+\to\phi\pi^+).\label{eq:inclu}
\end{equation}
In Fig.~\ref{inclu} the inclusive $\phi$ yield measured by CLEO-c
is plotted as a horizontal line and ${\cal{B}}_{\rm SUM}(D_S^+\to
\phi X)$ from Eq.~\ref{eq:inclu} is plotted as a function of
${\cal{B}}^{\rm eff}(D_S^+\to\phi\pi^+)$. The intersection point
gives the expected value of ${\cal{B}}^{\rm
eff}(D_S^+\to\phi\pi^+)$, if all the decay modes containing
$\phi$'s have been measured. More modes would increase the
expected inclusive rate and the shaded bands would rotate toward
the y axis. Thus, this technique gives an upper limit on
${\cal{B}}^{\rm eff}(D_S^+\to\phi\pi^+)$. Also shown are the
results for the $\eta'$ and $\eta$ modes \cite{eta-etap}.
 Averaging the intersection points gives
(3.25$\pm$0.46)\%, a value that is only meaningful as an upper
limit, i. e.
\begin{equation}
{\cal{B}}^{\rm eff}(D_S^+\to\phi\pi^+)< 3.85\%{\rm
~~(at~90\%~C.~L.)}.
\end{equation}
 It is clear that the inclusive measurements favor values for
${\cal{B}}^{\rm eff}(D_S^+\to\phi\pi^+)$ that are consistent with
the 3.5\% I have derived and are somewhat inconsistent with 4.8\%.

\subsection{The Real \boldmath{$(D_S^+\to\phi\pi^+)$} Branching Ratio}
In order to compare with theoretical calculations it is useful to
extract the value of the real $D_S^+\to\phi\pi^+$ branching ratio.
This can be done by using the results of a fit to the $K^-K^+\pi^+$
Dalitz plot (e. g. FOCUS) to get the fraction of $\phi\pi^+$. This
is not the same procedure that was done in the past of merely
cutting on the $K^+K^-$ invariant mass about the $\phi$.

The FOCUS Dalitz plot analysis determines the $\phi\pi^+$ fraction
as 0.45$\pm$0.01 \cite{Focus}. Multiplying the CLEO number for
${\cal{B}}(D_S^+\to K^+K^-\pi^+)$ by this $\phi\pi^+$ fraction
(and dividing by ${\cal{B}}(\phi\to K^+K^-)$ of 0.491), gives
\begin{equation}
{\cal{B}}(D_S^+\to\phi\pi^+)=(4.16\pm 0.41)\%~.
\end{equation}

\subsection{New Results on Singly and Doubly Cabibbo Suppressed
Decays}

\begin{figure}[h]
\centering
\includegraphics[width=80mm]{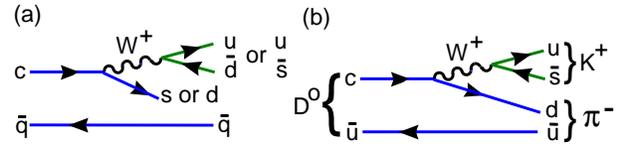}
\caption{(a) A spectator diagram for $D$ meson decays. If $c\to d$
or $W^+\to u\overline{s}$, the decay is said to be ``singly-Cabibbo
suppressed." If both occur in the same decay, we have a
``doubly-Cabibbo suppressed" decay. (b) Diagram for the
doubly-Cabibbo suppressed decay $D^0\to K^+\pi^-$.}
\label{Cabibbo-sup}
\end{figure}

In charm decays the $c$ quark can decay into an $s$ quark or a $d$
quark and a virtual $W^+$ boson. The decays into an $s$ quark are
Cabibbo allowed, since their rate is proportional to the CKM
element $|V_{cs}|^2$, while those into a $d$ quark are Cabibbo
suppressed, since their rate is proportional to $|V_{cd}|^2$,
where $|V_{cs}|\sim 0.97$, and $|V_{cd}|\sim 0.23$ \cite{PDG}. In
addition the virtual $W^+$ can form a Cabibbo favored
$u\overline{d}$ pair or a Cabibbo suppressed $u\overline{s}$ pair,
as shown in Fig.~\ref{Cabibbo-sup}(a). If $c\to d$ and $W^+\to
u\overline{s}$ we have a ``doubly-Cabibbo suppressed" decay, as
illustrated for one channel in Fig.~\ref{Cabibbo-sup}(b). CLEO-c
has measured a plethora of new singly-Cabibbo suppressed modes
\cite{cleo-sc}. Their measurements are listed in
Table~\ref{tab:tabsc}.

\begin{table}[hbt]
\begin{center}
\caption{Branching fractions for singly-Cabibbo suppressed $D$
decays from CLEO-c. The sources of the listed uncertainties are
statistical, experimental systematic, normalization mode, and  $CP$
correlations (for $D^0$ modes only). \label{tab:tabsc}}
\begin{tabular}{|l|c|c|}\hline
\boldmath{$D^0$}\textbf{ Modes}        & \boldmath{${\cal{B}}$
($10^{-3}$)}
\\\hline
\hline $\pi^+\pi^-$ & $~~~1.39\pm0.04\pm0.04\pm0.03\pm0.01~~~~$
\\\hline
$\pi^0\pi^0$ & $0.79\pm0.05\pm0.06\pm0.01\pm0.01$  \\\hline
$\pi^+\pi^-\pi^0$ & $13.2\pm0.2\pm0.5\pm0.2\pm0.1$
\\\hline $\pi^+\pi^+\pi^-\pi^-$  &
$7.3\pm0.1\pm0.3\pm0.1\pm0.1$  \\\hline $\pi^+\pi^-\pi^0\pi^0$ &
 $9.9\pm0.6\pm0.7\pm0.2\pm0.1$
\\\hline $\pi^+\pi^+\pi^-\pi^-\pi^0$  & $4.1\pm0.5\pm0.2\pm0.1\pm0.0$
\\\hline $\omega\pi^+\pi^-$ &
$1.7\pm0.5\pm0.2\pm0.0\pm0.0$   \\\hline $\eta\pi^0$  &
 $0.62\pm0.14\pm0.05\pm0.01\pm0.01$
\\\hline
$\pi^0\pi^0\pi^0$ & $<0.35$ (90\% CL)   \\\hline $\omega\pi^0$ &
$<0.26$ (90\% CL)    \\\hline $\eta\pi^+\pi^-$ &   $<1.9$ (90\%
CL)    \\\hline \hline \boldmath{$D^+$ }\textbf{Modes}        &
 \\\hline $\pi^+\pi^0$  &
1.25$\pm$0.06$\pm$0.07$\pm$0.04
 \\\hline $\pi^+\pi^+\pi^-$ &
 3.35$\pm$0.10$\pm$0.16$\pm$0.12  \\\hline $\pi^+\pi^0\pi^0$ &
4.8$\pm$0.3$\pm$0.3$\pm$0.2   \\\hline $\pi^+\pi^+\pi^-\pi^0$  &
11.6$\pm$0.4$\pm$0.6$\pm$0.4
\\\hline $\pi^+\pi^+\pi^+\pi^-\pi^-$ &
1.60$\pm$0.18$\pm$0.16$\pm$0.06 \\\hline $\eta\pi^+$ &
3.61$\pm$0.25$\pm$0.23$\pm$0.12
\\\hline $\omega\pi^+$ &   $<0.34$ (90\% CL)    \\\hline
\end{tabular}
\end{center}
\end{table}

There are also two new measurements from BaBar \cite{Babar-private}
one in the singly-Cabibbo suppressed decay mode
\begin{equation}
{\cal{B}}(D^+\to\pi^+\pi^0)= (1.22 \pm 0.10\pm 0.08 \pm
0.08)\times 10^{-3},
\end{equation}
which agrees with the CLEO-c result, and one in the doubly-Cabibbo
suppressed decay mode
\begin{equation}
{\cal{B}}(D^+\to K^+\pi^0)= (0.246 \pm 0.046\pm 0.024 \pm
0.016)\times 10^{-3},
\end{equation}
which is the first observation of this mode. Here BaBar normalizes
to the PDG absolute branching ratios \cite{PDG}. Normalizing to the
new world average numbers would substantially reduce the last error.

The di-pion system can have a total isospin of 2, or 0. The decays
into the three final states $\pi^+\pi^0$, $\pi^+\pi^-$ and
$\pi^0\pi^0$ starting from the $I=1/2$ $D$ meson states proceed
through a combination of $I=1/2$ and $I=3/2$ amplitudes
\cite{isospin}.
  These new $D\to\pi\pi$ branching fractions allow for a much more
  precise determination of the phase difference between these two
  isospin amplitudes.
The ratio of the $\Delta I=3/2$ to $\Delta I=1/2$ isospin
amplitudes and their relative strong phase difference is
$A_2/A_0=0.420\pm0.014\pm0.01$ and
$\cos\delta_I=0.062\pm0.048\pm0.058$ using the CLEO-c results
only. The large phase shift,
$\delta_I=(86.4\pm2.8\pm3.3)^{\circ}$, shows that final state
interactions are important in $D\to\pi\pi$ transitions. This
information could be useful in the study of $B\to\pi\pi$ decays
\cite{guo}.

\section{Searches for New Physics in Charm Decays}

\begin{figure}[h]
\centering
\includegraphics[width=80mm]{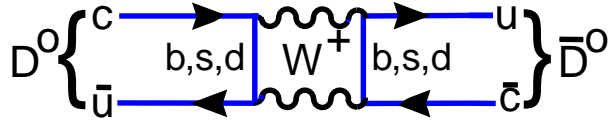}
\caption{A diagram for $D^0-\overline{D}^0$ mixing in the Standard
Model.} \label{Cmix2}
\end{figure}

New physics could be seen in charm decays by observations of mixing,
CP violation \cite{Bigi} or even T violation \cite{TBigi}. Let us
first consider $D^0-\overline{D}^0$ mixing. In the SM mixing is
generated by short distance diagrams including the one shown in
Fig.~\ref{Cmix2}. Here the heaviest intermediate quark is the $b$.
Since the mixing rate goes as the square of the mass of the
intermediate quark, we can see why it is suppressed relative to
$K^0$ mixing (50\%) or $B^0$ mixing (20\%), since the top-quark is
active in these systems. The CKM couplings also matter. That is why
$B_S$ mixing (50\%) is larger than $B^0$ mixing. For $D^0$ mixing
via the $b$-quark the couplings are $|V_{ub}|$ and $|V_{cb}|$, which
are also small.

\begin{figure}[h]
\centering
\includegraphics[width=80mm]{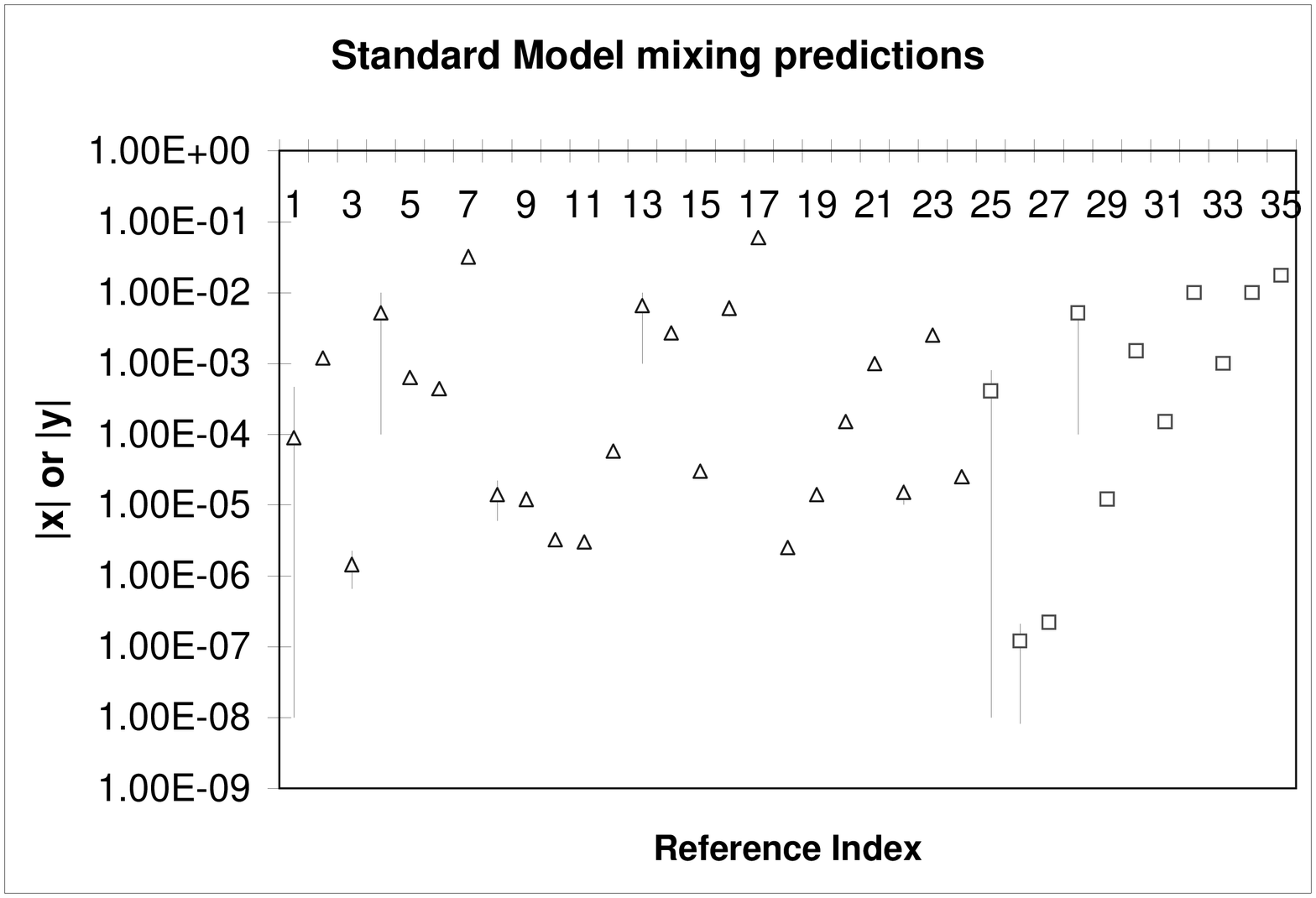}
\includegraphics[width=80mm]{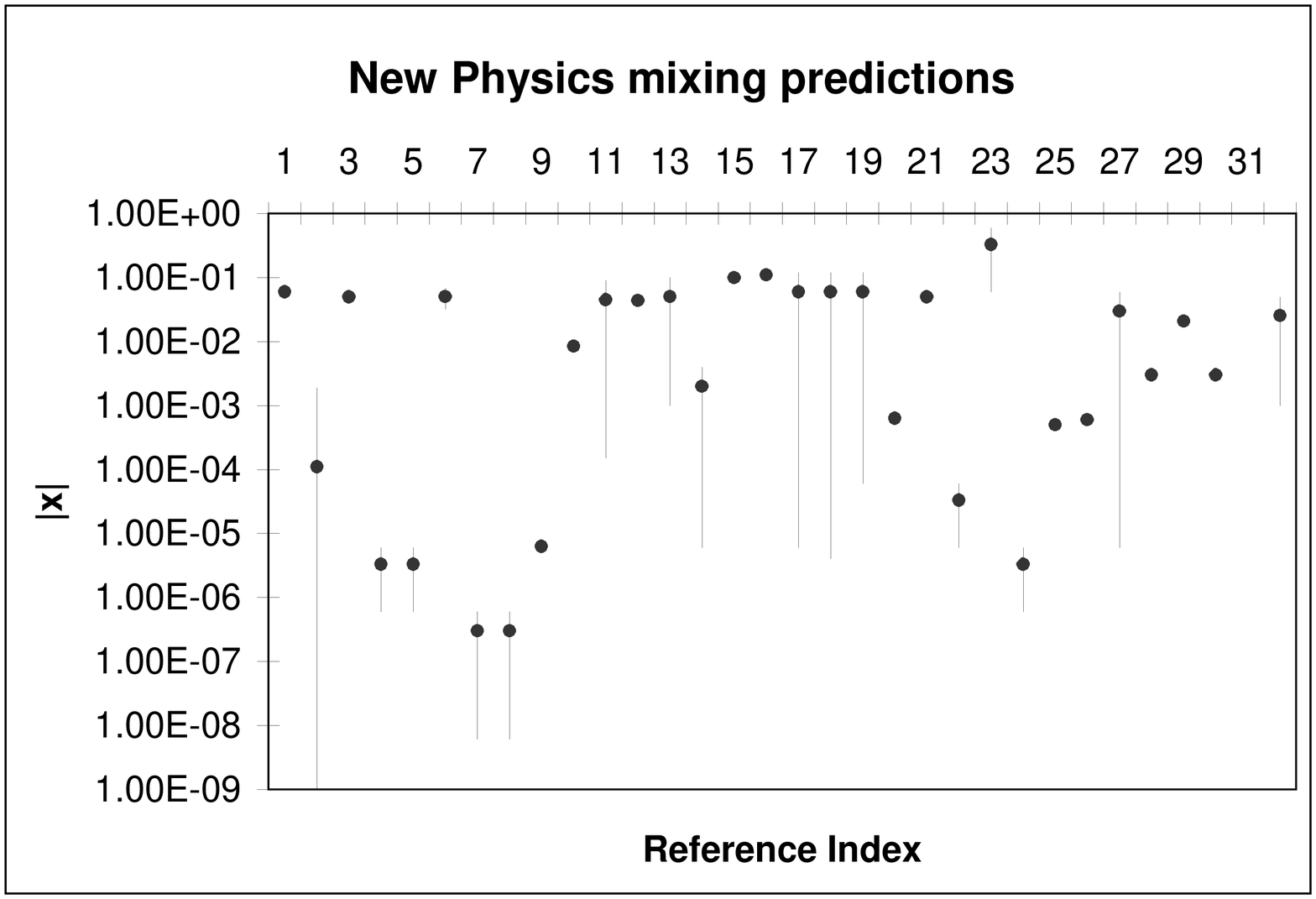}
\caption{Predictions for $x$ and $y$ in the Standard Model (top)
and for $x$ in New Physics models (bottom) indexed as to reference
number given in \cite{Petrov}.} \label{mixing-petrov}
\end{figure}

Mixing due to natural causes in the SM can be enhanced by so called
``long distance" effects, which are more-or-less the transition of a
$D^0$ into an on-shell meson pair, e.g. $K^+K^-$ and then another
transition back to a $\overline{D}^0$. Mixing is characterized by
the mass difference, $\Delta m$, and width difference
$\Delta\Gamma$, between CP+ and CP- eigenstates, where the width
$\Gamma$ is related to the lifetime, $\tau_{D^0}$, as
$\Gamma\cdot\tau_{D^0}=\hbar$. New Physics effects in loops, for
example new particles, would tend to cause $x\equiv \Delta m/\Gamma
>> y\equiv \Delta\Gamma/2\Gamma$. Predictions though of the
magnitude of SM and NP effects are murky. Fig.~\ref{mixing-petrov}
from Petrov, an update to a plot originally shown by Nelson, shows
various expectations \cite{Petrov}.

Both SM and NP predictions range over the entire plot which will
make it difficult to claim a NP effect. Nevertheless, it will be
interesting to see or limit the size of $x$ and $y$, especially
since tight limits will constrain NP models, and observable mixing,
whatever the source, could be a path for NP to generate CP
violation.

\subsection{\boldmath{$D^0\overline{D}^0$} Mixing Using Wrong-Sign \boldmath{$K^-\pi^+$} Decay}
\begin{figure}[ht]
\centering
\includegraphics[width=76mm]{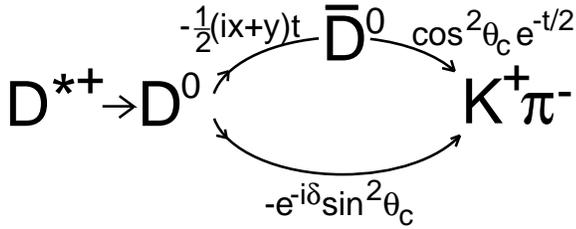}
\vspace{0.1cm} \caption{Amplitudes for $D^0$ decay into $K^+\pi^-$
either via mixing followed by a Cabibbo favored decay (top path), or
doubly-Cabibbo suppressed decay (bottom path).} \label{mixKpi}
\end{figure}

The $D^{*\pm}\to \pi^{\pm} D^0$ decay provides a useful tag of the
initial quark content, i. e., whether we start with a $D^0$ or a
$\overline{D}^0$. The $\pi^+$ tags $D^0$, and the $\pi^-$ tags
$\overline{D}^0$. While the normal Cabibbo favored decay is
$D^0\to K^-\pi^+$, it is possible to get $D^0\to K^+\pi^-$, a
``wrong-sign" decay, via mixing. Unfortunately, this clean
signature is complicated by doubly-Cabibbo suppressed decays, as
shown in Fig.~\ref{Cabibbo-sup}(b). The two processes interfere.
Fig.~\ref{mixKpi} diagrammatically shows the two decay paths.

The interference causes the measured $x$ and $y$ to be rotated
through an angle $\delta$, the phase difference between the
doubly-Cabibbo suppressed and Cabibbo favored processes. The
wrong-sign decay rate then as a function of time is
\begin{equation}
R_{\rm ws}=e^{-\Gamma t}\left(R_D+\sqrt{R_D}y'\Gamma
t+\frac{1}{4}\left(x'^2+y'^2\right)(\Gamma t)^2\right)~,
\end{equation}
where $R_D$ is the doubly-Cabibbo suppressed decay rate and $\Gamma$
is the decay width. By measuring the time dependence of the decay
rate it is possible to sort out the mixing from the doubly-Cabibbo
suppressed decay. Several measurements have been made resulting in
the upper limits listed in Table~\ref{tab:mixKpi} on $x'^2$ and
$y'$.

\begin{table}[hbt]
\begin{center}
\caption{Limits on $x'^2$ and $y'$, both at 95\% C. L. for cases
where CP violation is allowed and also where it is not allowed.
(Note that the limits are on $x'^2$ not $x'$) . \label{tab:mixKpi}}
\begin{tabular}{|l|c|c|c|c|}\hline
\textbf{Exp.}        &
\multicolumn{2}{c}{\boldmath{$x'^2(\times10^{-3}$)}} &
\multicolumn{2}{|c|}{\boldmath{$y'(\times10^{-3}$)}}\\
& \textbf{CPV} &\textbf{No CPV} & \textbf{CPV} &\textbf{No CPV}
\\\hline
CLEO\cite{kpi-cleo} & 0.82 & 0.78 & -58$<y'<$10 & -52$<y'<$2\\\hline

FOCUS\cite{kpi-focus}& 0.80 & 0.83 & -120$<y'<$67 &
-72$<y'<$41\\\hline

Belle\cite{kpi-belle}& 0.72 & 0.72 & -28$<y'<$21 &
-9.9$<y'<$6.8\\\hline

BaBar\cite{kpi-babar}& 2.2 & 2.0 & -56$<y'<$39 & -27$<y'<$22\\\hline
\end{tabular}
\end{center}
\end{table}

The limits are somewhat more restrictive on $y'$ when CP violation
is not permitted, while those on $x'^2$ hardly change. While no
experiment claims an effect, it is interesting that the Belle result
is consistent with no mixing only at 3.9\% C. L. \cite{kpi-belle}.

\subsection{Other Mixing Studies}
There are several other methods that have been used to search for
mixing. The lifetime difference between mass eigenstates provides
another measurement of $y$, usually called $y_{CP}$. In presence of
$CP$ violation, $y_{CP}$ is a linear combination of $x$ and $y$
involving the $CP$ violation phase $\phi$. Table~\ref{sum-ycp}
summarizes experimental data on $y_{CP}$. The average is
0.90$\pm$0.42, which is consistent with zero.

\begin{table}[htb]
\caption{Summary of $y_{CP}$ results.}\label{sum-ycp}
\begin{center}
\begin{tabular}{|l|c|}
\hline \textbf{Experiment} &
\boldmath{$y_{CP}$}\textbf{(\%)}\\
\hline E791\cite{ycp-e791} & $0.8\pm 2.9\pm 1.0$\\
\hline FOCUS \cite{ycp-focus} & $3.4\pm 1.4 \pm 0.7$\\\hline CLEO
\cite{ycp-cleo} &  $-1.2\pm 2.5\pm 1.4$\\\hline Belle, untagged
\cite{belle-untg}&$-0.5\pm 1.0\pm0.8$\\\hline Belle, tagged
\cite{belle-tag}&$1.2\pm 0.7\pm 0.4$\\\hline BaBar \cite{ycp-babar}&
$0.8\pm 0.4 ^{+0.5}_{-0.4}$\\\hline
\end{tabular}
\end{center}
\end{table}

Mixing measurements have also been made in semileptonic decay modes
where doubly-Cabibbo suppressed decays are absent. Here the mixed
rate $R_M=(x^2+y^2)/2$ is directly measured.  Table~\ref{dmix:sl}
summarizes the present experimental limits.

\begin{table}[htb]
\caption{Summary of mixing limits (90 \% C.L.) from $D^0$
semileptonic decay studies.}\label{dmix:sl}
\begin{center}
\begin{tabular}{|l|c|c|}
\hline \textbf{Experiment} & \boldmath{$R_M$}&
\boldmath{$\sqrt{x^2+y^2}$}\\\hline FOCUS \cite{sl-cleo} & 0.0078 &
0.12
\\\hline BaBar \cite{sl-babar} & 0.0042 & 0.096
\\\hline Belle \cite{sl-belle} & 0.0010 & 0.044 \\\hline
\end{tabular}
\end{center}
\end{table}

\begin{figure}[ht]
\centering
\includegraphics[width=80mm]{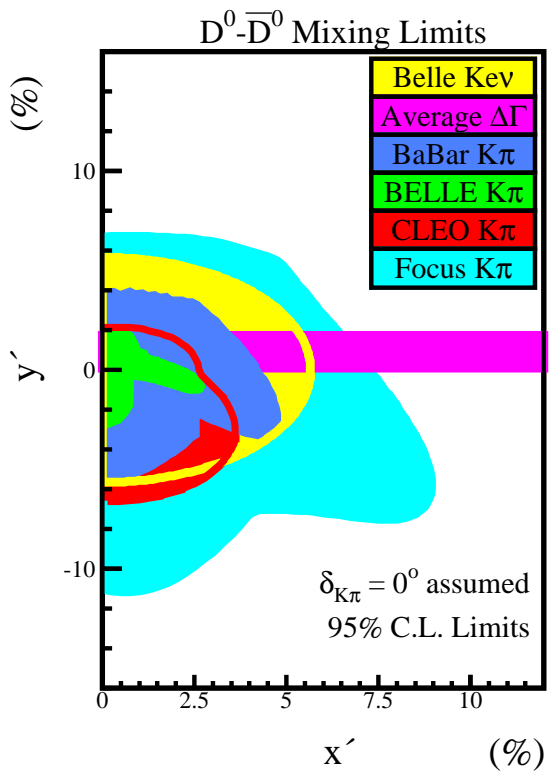}
\vspace{-3 mm}\caption{Limits at 95\% C. L. on $x'$ and $y'$ from
various measurements described in the text. Here $\Delta\Gamma$
refers to the width difference between CP+ and CP- eigenstates found
by measuring the lifetime difference.} \label{pdg2006}
\end{figure}

Fig.~\ref{pdg2006}, prepared by D.~Asner for an upcoming PDG review
summarizes the overall situation on $D^0-\overline{D}^0$ mixing from
the above measurements.

Another way of searching for mixing is to use Dalitz plot analyses
of three-body decay modes. CLEO has done a full time dependent
analysis of the $D^0\to K_S\pi^+\pi^-$ mode \cite{cleo-ks2pib}. The
essential feature here is that the CP+ $D_1$ state has a different
time dependence than the CP- $D_2$ state
\begin{eqnarray}
D_1(t)&\sim&
\exp\left[-i\left(m_1-i\Gamma_1/2\right)t\right]\\\nonumber
D_2(t)&\sim&
\exp\left[-i\left(m_2-i\Gamma_2/2\right)t\right].\\\nonumber
\end{eqnarray}

Limits extracted are  $(-4.5<x<9.3)$\% and $(-6.4<y<3.6)$\% at 95\%
C.L., without assumptions regarding CP-violating parameters. This
result is compared with others in Fig.~\ref{pdg2006} (again from
D.~Asner). We note that the CLEO limits are comparable even though
they are based on an order of magnitude less luminosity, showing the
potential of such analyses.

\begin{figure}[ht]
\centering
\includegraphics[width=80mm]{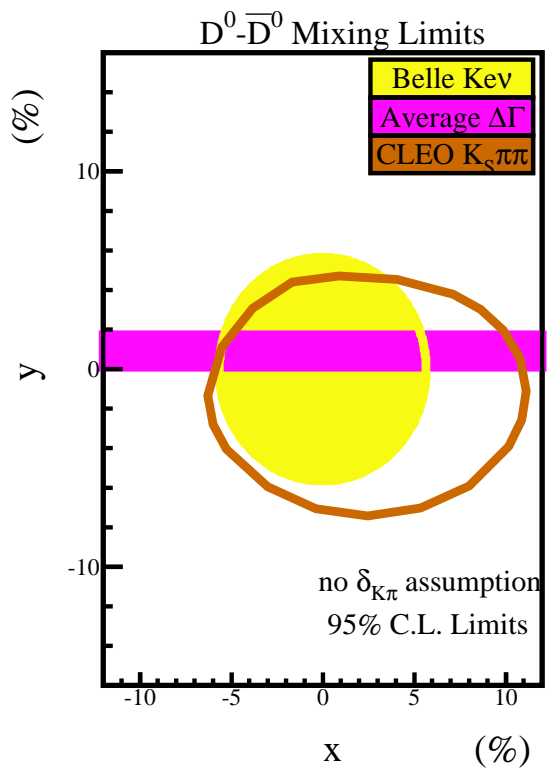}
\vspace{-5mm} \caption{Limits at 95\% C. L. on $x$ and $y$ from the
CLEO $K_S\pi^+\pi^-$ Dalitz analysis, the average lifetime
difference and the Belle semileptonic analysis.} \label{pdg2006b}
\end{figure}

BaBar at this conference presented an analysis of $D^0\to
K^+\pi^-\pi^0$ where they use cuts on the Dalitz plot to enhance the
Cabibbo favored rate and suppressed the doubly-Cabibbo suppressed
rate \cite{Wilson}. This is possible since the Cabibbo favored rate
proceeds largely through $K^-\rho^+$, while the doubly-Cabibbo
suppressed rate goes into $K^{*+}\pi^-$ and $K^{*0}\pi^0$. For the
CP conserving fit they find $R_M=(0.23^{+0.18}_{-0.14}\pm
0.04)\times 10^{-3}$ which translates into $R_M < 0.54\times
10^{-3}$ at 95\% C. L.. They also find that $R_M$ is consistent with
no mixing at 4.5\% C. L..

\subsection{CP/T Violation}

Unexpectedly large CP violating asymmetries in the range from
$10^{-2}-10^{-4}$ is a better signature for NP than mixing.  There
are several ways to study CP violation in charm decays
\cite{Apetrov}. We can look for direct CP violation, even in charged
decays \cite{Direct}; we can look for CP violation via mixing; T
violation can be examined in 4-body $D$ meson decays, assuming CPT
conservation, by measuring triple-product correlations \cite{TBigi}.
Finally the quantum coherence present in correlated
$D^0-\overline{D^0}$ decays of the $\psi(3770)$ can be exploited
(see the talk by D. Cinabro \cite{Cinabro}). Some recent results are
shown in Table~\ref{tab:CP}. No significant effects have been seen.

\begin{table}[h]
\begin{center}
\caption{Measurements of CP violating asymmetries, $A_{CP}$ ( or
$A_{T}$) in charm decays. BaBar compares $D^+$ versus $D^-$ rates.
The CLEO result is obtained using a Dalitz plot analysis. CDF uses
the decay of $D^{*+}\to \pi^+ D^0$ as a flavor tag, and FOCUS uses
triple product correlations to measure T violation. }
\begin{tabular}{|l|l|c|}
\hline \textbf{Exp.} & \textbf{Decay Mode} & \boldmath{$A_{CP}$ or
$A_T$}\textbf{(\%)}
\\\hline

BaBar \cite{babar-cpv} & $D^+\to K^-K^+\pi^+$ & $1.4\pm1.0\pm 0.8$
\\\hline
BaBar \cite{babar-cpv}& $D^+\to \phi\pi^+$ & $0.2\pm1.5\pm 0.6$
\\\hline
BaBar \cite{babar-cpv}& $D^+\to K^{*0} K^+$ & $0.2\pm1.5\pm 0.6$
\\\hline

Belle \cite{belle-cpv}& $D^0\rightarrow K^+\pi^-\pi^0$ & -0.6$\pm$
5.3\\\hline Belle \cite{belle-cpv}& $D^0\rightarrow
K^+\pi^-\pi^+\pi^-$ & $-1.8\pm 4.4$\\\hline

CLEO \cite{cleo-cpv}& $D^0\to\pi^+\pi^-\pi^0$ & $1^{+9}_{-7}\pm 8$
\\\hline

CLEO \cite{cleo-cpv2}& $D^0\to K^-\pi^+\pi^0$ & $-3.1\pm 8.6$
\\\hline
CLEO \cite{cleo-cpv3}& $D^0\to K_S\pi^+\pi^-$ & $-0.9\pm
2.1^{+1.0+1.3}_{-4.3-3.7}$
\\\hline

CDF \cite{CDF-cpv} &$D^0\to K^+ K^-$& $2.0\pm 1.2 \pm 0.6$
\\\hline
CDF \cite{CDF-cpv}&$D^0\to\pi^+\pi^-$ &$1.0\pm 1.3\pm 0.6$
\\\hline
FOCUS \cite{FOCUS-cpv} & $D^0\to K^+K^-\pi^+\pi^-$ & $1.0\pm 5.7\pm
3.7$\\\hline

FOCUS \cite{FOCUS-cpv} & $D^+\to K^0K^+\pi^+\pi^-$ & $2.3\pm 6.2\pm
2.2$\\\hline

FOCUS \cite{FOCUS-cpv} & $D_S^+\to K^0K^+\pi^+\pi^-$ & $-3.6\pm
6.7\pm 2.3$\\\hline
\end{tabular}
\label{tab:CP}
\end{center}
\end{table}

\subsection{Conclusions}

The absolute branching fractions for charm mesons have been measured
with unprecedented accuracy. Combining the PDG values with the
preliminary CLEO-c results for $D^0$ and $D^+$ decays, and using the
CLEO-c results for $D_S^+$, I find
\begin{eqnarray}
{\cal {B}}(D^0\to K^-\pi^+)&=& (3.87\pm 0.06)\% \\
{\cal {B}}(D^+\to K^-\pi^+\pi^+)&=& (9.12\pm 0.19)\%\\
{\cal {B}}(D_S^+\to K^-K^+\pi^+)&=&(4.54^{+0.44}_{-0.42}\pm
0.25)\%.\\\nonumber
\end{eqnarray}

CLEO-c does not quote a branching ratio for $D_S^+\to\phi\pi^+$
mode because of interferences on the Dalitz plot. The
$K^-K^+\pi^+$ or the $K^0K^+$ modes should be used for
normalization. Since most of the $D_S$ decay modes have been
measured as ratios to the $\phi\pi^+$ mode, I extract an effective
branching ratio
\begin{equation}
{\cal{B}}^{\rm eff}(D_S^+\to\phi\pi^+)=(3.49\pm 0.39)\%.
\end{equation}

These rates can be used for many purposes. For example, adding up
the number of charm quarks produced in each $B$ meson decay at the
$\Upsilon(4S)$ resonance by summing the $D^0$, $D^+$, $D_S^+$,
charmed baryon and twice the charmonium yields gives a rate of
1.09$\pm$0.04, where the largest error comes from the $D^0$ yield.

Many more Cabibbo suppressed and some doubly-Cabibbo suppressed
modes have been measured. Large phase shifts have been more
accurately measured in the $D\to\pi\pi$ channel.

There is no definitive evidence for $D^0-\overline{D^0}$ mixing. The
best limits yet are $|y'|<2.5$\%  and $|x'^2|~<~7.2\times 10^{-3}$
both at 95\% C. L. The limit on $|x'|$ of about 8\% is just
beginning to probe an interesting range. There are two hints that
mixing may be soon found. Belle finds consistency with no mixing at
3.9\% C. L. in wrong-sign $K^-\pi^+$ decays and BaBar finds
consistency with no mixing at 4.5\% C. L. in wrong sign
$K^-\pi^+\pi^0$ decays, thus making further searches more
interesting. There have not been any observations of CP or T
violation.

% If you have acknowledgments, this puts in the proper section head.
\bigskip % extra skip inserted
\begin{acknowledgments}
This work was supported by the National Science Foundation under
grant \#0553004. I thank M.~Artuso, D.~Asner, R.~Faccini,
S.~Malvezzi, N.~Menaa, P.~Onysi, R.~Sia and S.~Stroiney for
interesting discussions and providing data and plots used in this
review.
\end{acknowledgments}

\bigskip % extra skip inserted
% Create the reference section using BibTeX:
%\bibliography{basename of .bib file}

\begin{thebibliography}{9}   % Use for  1-9  references
%\begin{thebibliography}{99} % Use for 10-99 references
\bibitem{Bigi}
I. I. Bigi and A. I. Sanda, ``CP Violation,"  Cambridge Univ. Press,
Cambridge (2000) p. 252.

\bibitem{Artuso}
M. Artuso, ``Charm Decays Within the Standard Model and Beyond,"
in {\it Proc. of the XXII Int. Symp. on Lepton \& Photon
Interactions at High Energies}, ed. R. Brenner, C. P. de los
Heros, and J. Rathsman, World Scientific, Singapore (2006)
[hep-ex/0510052.

\bibitem{Poling}
R. Poling, ``CLEO-c Hot Topics," presented at Flavor Physics and CP
Violation, April 9-12, 2006, Vancouver, B. C., Canada.


\bibitem{cleo-dbr}
Q. He \etal~(CLEO), Phys. Rev. Lett. {\bf 95} 121801 (2005)
[hep-ex/0504003].

\bibitem{Dbr}
S. Stroiney, ``$D$ Hadronic Branching Fractions," presented at APS
April Meeting, April 22-25, 2006, Dallas, Texas.

\bibitem{PDG}
S. Eidelman \etal~(PDG), Phys. Lett. {\bf B592}, 1 (2004) and
http://pdg.lbl.gov/~.

\bibitem{Focus}
S. Malvezzi, ``$D$-Meson Dalitz Fit From Focus," Prepared for 7th
Conference on Intersections Between Particle and Nuclear Physics
(CIPANP 2000), Quebec City, Quebec, Canada, 22-28 May 2000.
Published in AIP Conf. Proc. {\bf 549}, 569 (2002).

\bibitem{E687}
P. L. Frabetti \etal.)~(E687), Phys. Lett. {\bf B351}, 591 (1995).

\bibitem{CLEO-DS1}
M. Artuso \etal ~(CLEO) Phys. Lett. {\bf B378} 364, (1996).

\bibitem{BaBar-DS}
B. Aubert \etal ~(BaBar), Phys. Rev. {\bf D71}, 091104 (2005)
[hep-ex/0502041].

\bibitem{Marsiske}
H. Marsiske, ``Charmonium/Charm Spectroscopy from B-Factories,"
presented at Flavor Physics and CP Violation, April 9-12, 2006,
Vancouver, B. C., Canada.

\bibitem{Muheim}
F. Muheim and S. Stone, Phys. Rev. {\bf D49}, 3767 (1994).

\bibitem{BES}
J. Z. Bai \etal~(BES), Phys. Rev. {\bf D52}, 3781 (1995).

\bibitem{eta-etap}
The $\eta'$ modes include $\eta'\pi^+$, $\eta'\pi^+\pi^0$, and
$\eta'\ell^+\nu$, that sum to 4.3$\pm$0.44 times ${\cal{B}}^{\rm
eff}(D_S^+\to\phi\pi^+)$. The $\eta$ modes include the analgous
modes to those for the $\eta'$ and, inaddition, feeddown from the
$\eta'$ modes; they sum to 9.93$\pm$0.95 times ${\cal{B}}^{\rm
eff}(D_S^+\to\phi\pi^+)$.

\bibitem{cleo-sc}
P. Rubin \etal~(CLEO) Phys. Rev. Lett. {\bf 96}, 08180 (2006)
[hep-ex/0512063].

\bibitem{Babar-private} Private communication from R. Faccini of the BaBar
collaboration.

\bibitem{isospin}
%H. Lipkin {\it et al.,} Phys. Rev. D {\bf 44}, 1454 (1991).
M. Selen {\it et al.} (CLEO), Phys. Rev. Lett. {\bf 71}, 1973
(1993).

\bibitem{guo}
P. Guo, X-G. He, and X-Q Li, Int. J .Mod. Phys. {\bf A21}, 57
(2006) [hep-ph/0402262]


\bibitem{TBigi}I.I. Bigi, in {KAON2001: International
Conference on CP Violation}, (2001) [hep-ph/0107102].


\bibitem{Petrov}
A. Petrov, ``Charm Physics: Theoretical Review,"  invited talk at
Flavor Physics and CP Violation (FPCP 2003), Paris, France, 3-6
Jun 2003; Published in eConf C030603: MEc05 (2003)
http://www.slac.stanford.edu/econf/C030603/\newline
[hep-ph/0311371].

\bibitem{kpi-cleo}R. Godang \etal\ (CLEO), Phys.\
Rev.\ Lett. {\bf 84}, 5038 (2000) [hep-ex/0001060].
%%CITATION = HEP-EX 0105002;%%
\bibitem{kpi-focus}
J.~M.~Link {\it et al.}  (FOCUS),
%``Measurement of the doubly Cabibbo suppressed decay D0 $\to$ K+ pi- and a
%search for charm mixing,''
Phys.\ Lett.\ B {\bf 618}, 23 (2005) hep-ex/0412034].
\bibitem{kpi-belle}
K.~Abe {\it et al.}  (BELLE),
%``Search for D0 anti-D0 mixing in D0 $\to$ K+ pi- decays and measurement of
%the doubly-Cabibbo-suppressed decay rate,''
Phys.\ Rev.\ Lett.\  {\bf 94}, 071801 (2005) hep-ex/0408125].
\bibitem{kpi-babar}B.~Aubert {\it et al.}  (BABAR),
%``Search for D0 - anti-D0 mixing and a measurement of the doubly
%Cabibbo-suppressed decay rate in D0 $\to$ K pi decays,''
Phys.\ Rev.\ Lett.\  {\bf 91}, 171801 (2003) [hep-ex/0304007].
%%CITATION = HEP-EX 0304007;%%

\bibitem{ycp-e791}
E. M. Aitala \etal~(E791), Phys. Rev. Lett. {\bf 83}, 32 (1999)
[hep-ex/9903012].

\bibitem{ycp-focus}J. Link {\em et al.} (FOCUS), Phys. Lett. B {\bf
485}, 62 (2000).
\bibitem{ycp-cleo}S.E. Csorna {\em et al.} (CLEO),
Phys. Rev. D {\bf 65}, 092001 (2002).
\bibitem{belle-untg}K. Abe {\em et al.} (Belle), Phys. Rev. Lett.
{\bf 88}, 162001 (2002).
\bibitem{belle-tag}K. Abe {\em et al.}, ``Meaurement of the
$D^0-\overline{D^0}$ Lifetime Difference Using $D^0\to K\pi/KK$
Decays," submitted to Lepton-Photon Conference [BELLE-CONF-347]
(2003).

\bibitem{ycp-babar}B. Aubert {\em et al.} (BaBar),
Phys. Rev. Lett. {\bf 91}, 171801 (2003).

\bibitem{sl-cleo}
C.~Cawlfield {\it et al.}  (CLEO),
%``Limits on neutral D mixing in semileptonic decays,''
Phys.\ Rev.\ D {\bf 71}, 077101 (2005) [hep-ex/0502012].
%%CITATION = HEP-EX 0502012;%%
\bibitem{sl-babar}
B.~Aubert {\it et al.}  (BABAR),
%``Search for D0 - anti-D0 mixing using semileptonic decay modes,''
Phys.\ Rev.\ D {\bf 70}, 091102 (2004) [hep-ex/0408066].
%%CITATION = HEP-EX 0408066;%%
\bibitem{sl-belle}U. Bitenc {\it et al.}  (Belle), Phys. Rev. {\bf D72}, 071101 (2005)
%``Search for D0 anti-D0 mixing using semileptonic decays at Belle,''
[hep-ex/0507020].
%%CITATION = HEP-EX 0507020;%%

\bibitem{cleo-ks2pib}
D.~M.~Asner \etal~  (CLEO),
%``Search for anti-D0 D0 mixing in the Dalitz plot analysis of D0 $\to$ K0(S)
%pi+ pi-,''
Phys.\ Rev.\ D {\bf 72}, 012001 (2005) [hep-ex/0503045].

\bibitem{Wilson}
M. Wilson, ``BaBar Hot Topics (II),"  presented at Flavor Physics
and CP Violation, April 9-12, 2006, Vancouver, B. C., Canada.


\bibitem{Apetrov}A.~A.~Petrov,
%``Mixing and CP-violation in charm,''
Nucl.\ Phys.\ Proc.\ Suppl.\  {\bf 142}, 333 (2005)
[hep-ph/0409130].

\bibitem{Direct}
As usual it takes two interfering amplitudes to generate an
asymmetry, so no direct SM CP asymmetries can arise in pure Cabibbo
allowed or doubly-Cabibbo suppressed decays. See ref.~\cite{Bigi} p.
258. Asymmetries in singly-Cabibbo suppressed decays are very small,
on the order of $|V_{us}|^4$ in the SM.

\bibitem{Cinabro}
D. Cinabro, ``Interference Effects in D Meson Decays," presented
at Flavor Physics and CP Violation, April 9-12, 2006, Vancouver,
B. C., Canada. See also  W.~M.~Sun,
  %``D0 anti-D0 quantum correlations, mixing, and strong phases,''
[hep-ex/0603031].

%
\bibitem{babar-cpv}
B.~Aubert {\it et al.}  (BABAR),
%``A search for CP violation and a measurement of the relative branching
%fraction in D+ $\to$ K- K+ pi+ decays,''
Phys.\ Rev.\ D {\bf 71}, 091101 (2005) [hep-ex/0501075].
%%CITATION = HEP-EX 0501075;%%


\bibitem{belle-cpv}
X.~C.~Tian {\it et al.}  (Belle), Phys. Rev. Lett. {\bf 95},
231801 (2005)
%``Measurement of the wrong-sign decays D0 $\to$ K+ pi- (pi0, pi+ pi-) and
%search for CP violation,''
[hep-ex/0507071].
\bibitem{cleo-cpv}D.~Cronin-Hennessy {\it et al.}  (CLEO),
%``Searches for CP violation and pi pi S-wave in the Dalitz-plot of D0 $\to$ pi+
%pi- pi0,''
Phys.\ Rev.\ D {\bf 72}, 031102 (2005) [hep-ex/0503052].
%%CITATION = HEP-EX 0503052;%%
\bibitem{cleo-cpv2}
 S.~Kopp {\it et al.}  (CLEO),
  %``Dalitz analysis of the decay D0 $\to$ K- pi+ pi0,''
  Phys.\ Rev.\ D {\bf 63}, 092001 (2001)
  [hep-ex/0011065].

\bibitem{cleo-cpv3}
D.~M.~Asner {\it et al.}  (CLEO),
  %``Search for CP violation in D0 $\to$ K0(S) pi+ pi-,''
  Phys.\ Rev.\ D {\bf 70}, 091101 (2004)
  [hep-ex/0311033].

  \bibitem{CDF-cpv}
D. Acosta \etal~(CDF) Phys. Rev. Lett. {\bf 94}, 122001 (2005)
[hep-ex/0504006].


\bibitem{FOCUS-cpv}
J.~M.~Link {\it et al.}  (FOCUS),
%``Search for T violation in charm meson decays,''
Phys.\ Lett.\ B {\bf 622}, 239 (2005) [hep-ex/0506012].
%%CITATION = HEP-EX 0506012;%%
\end{thebibliography}

\end{document}